\documentclass[Crown, sagev, times]{sagej}
\usepackage{moreverb, url, dsfont, soul, dsfont, amsmath, tikz, natbib}
\usepackage{xr, xr-hyper}
\usepackage[colorlinks, bookmarksopen, bookmarksnumbered, citecolor=red, urlcolor=red]{hyperref}
\usepackage{subcaption, threeparttable, multicol, multirow, lscape, makecell, setspace}
\usepackage{enumitem}
\usepackage[title]{appendix}
\usepackage{mathtools}
\usepackage{geometry}
\usepackage{floatrow, inputenc}
\newcommand{\Ex}{\mathbb{E}}
\newcommand{\var}{\text{var}}
\newcommand{\cov}{\text{cov}}
\newcommand{\mc}{\mathcal}
\newcommand{\bd}{\boldsymbol}
\newtheorem{theorem}{Theorem}
\newcommand\BibTeX{{\rmfamily B\kern-.05em \textsc{i\kern-.025em b}\kern-.08em
		T\kern-.1667em\lower.7ex\hbox{E}\kern-.125emX}}

\def\volumeyear{2019}
\definecolor{jcolor}{RGB}{041,122,000}
\definecolor{darkred}{RGB}{100,000,000}
\definecolor{purple}{RGB}{200,000,200}

\begin{document}
	
\runninghead{Liu et al.}
\title{Average treatment effect on the treated, under lack of positivity} 
\author{Yi Liu\affilnum{1}, Huiyue Li\affilnum{2}, Yunji Zhou\affilnum{3}, and Roland A. Matsouaka\affilnum{4,}\affilnum{5} }
\affiliation{\affilnum{1} Department of Statistics, North Carolina State University, Raleigh, North Carolina\\
\affilnum{2} Baim Institute for Clinical Research, Boston, Massachusetts\\
\affilnum{3} Department of Biostatistics, University of Washington, Seattle, Washington\\
\affilnum{4} Department of Biostatistics and Bioinformatics, Duke University, Durham, North Carolina\\
\affilnum{5} Program for Comparative Effectiveness Methodology, Duke Clinical Research Institute,  Durham, North Carolina}
\corrauth{Roland A. Matsouaka, Duke Clinical Research Institute, 200  Morris St., Room 7826, Durham, NC 27701}
	
\email{roland.matsouaka@duke.edu}
	
\begin{abstract}\small

The use of propensity score (PS) methods has become ubiquitous in causal inference. At the heart of these methods is the positivity assumption. Violation of the positivity assumption leads to the presence of extreme PS weights when estimating average causal effects of interest, such as the average treatment effect (ATE) or the average treatment effect on the treated (ATT), which renders invalid related statistical inference. To circumvent this issue, trimming or truncating the extreme estimated PSs have been widely used. However, these methods require that we specify a priori a threshold and sometimes an additional smoothing parameter. While there are a number of methods dealing with the lack of positivity when estimating ATE, surprisingly there is no much effort in the same issue for ATT.  In this paper, we first review widely used methods, such as trimming and truncation in ATT. We emphasize the underlying intuition behind these methods to better understand their applications and highlight their main limitations.  
Then, we argue that the current methods simply target estimands that are scaled ATT (and thus move the goalpost to a different target of interest), where we specify the scale and the target populations. We further propose a PS weight-based alternative for the average causal effect on the treated, called overlap weighted average treatment effect on the treated (OWATT). The appeal of our proposed method lies in its ability to obtain similar or even better results than trimming and truncation while relaxing the constraint to choose a priori a threshold (or even specify a smoothing parameter). The performance of the proposed method is illustrated via a series of Monte Carlo simulations and a data analysis on racial disparities in health care expenditures. 
		
\end{abstract}

\maketitle	

%%%%%%%%%%%%%%%%%%%%%%%%%%%%%%%%%%% Start of the article %%%%%%%%%%%%%%%%%%%%%%%%%%%%%%%%%%%%%%%%%%%%%%%%%%%%%%%%%%%%%%%%%%%

\section{Introduction}\label{sec:intro}

The use of propensity score (PS) methods has become ubiquitous in causal inference from observational studies. It provides a proper framework to assess treatment effects while properly adjusting for confounding or selection bias. At the heart of PS methods is the positivity assumption, which specifies that each study participant has a non-zero probability to be assigned to all the treatment options, given their covariates.\citep{rosenbaum1983central} {Violation (or near violation) of the positivity assumption often leads to  estimated PSs that are near (or equal to) 0 or 1. These, in return, {can yield extreme PS weights.} %yield extreme PS weights when estimating (average) causal effects, rendering invalid inference or unstable asymptotic normal approximation. 
In other words, without the positivity assumption, it is often difficult to obtain adequate asymptotic results for the effect of interest.
\citep{busso2014new, ma2020robust, d2021overlap}}

To circumvent the issue caused by these %heavy
{large} weights, we often trim or truncate extreme estimated PS weights based on a user-specified threshold $\alpha\in(0,0.5)$. The value of the threshold $\alpha$ has to be small enough as to enclose a sufficient number of observations in the final sample, but large enough to avoid the undue influence of participants with extreme PS weights. For instance, to estimate the average treatment effect (ATE) via trimming, we often exclude participants whose PSs fall outside the interval $[0.1,0.9]$.\citep{cole2003effect, sturmer2006insights, kurth2005results, sturmer2010treatment, crump2006moving, dehejia1999causal, yang2018asymptotic} However, the selection of a reasonable threshold remains somewhat subjective.  Although Crump et al. \cite{crump2009dealing} proposed  a rule-of-thumb to select a trimming threshold selection, the  method is not always optimal as it depends on the size of the data at hand. \cite{zhou2020propensity} Furthermore, trimming can render the target estimand not root-$n$ estimable, i.e., the standard central limit theorem may not apply to draw inference on the treatment effect over the target population as the asymptotic variance may not be {bounded}. \citep{crump2006moving, crump2009dealing, yang2018asymptotic} {Besides trimming, one can also use truncation   (or weight capping) to deal with extreme weights. With truncation,  PSs outside pre-specified thresholds are replaced with the threshold values. One can truncate extreme weights, for instance, by replacing any PS that falls outside the interval [0.1,0.9]   with 0.9 (resp. 0.1) if the PS is greater than 0.9 (resp. lower than 0.1). Both truncation and trimming can be done via data-adaptive thresholds, e.g., using some percentiles of the PSs.\cite{sturmer2010treatment,ju2019adaptive,gruber2022data} }

As an alternative, Li et al.\cite{li2018balancing} proposed the use of overlap weights (OW) to target a different estimand, the average treatment effect on the overlap population (ATO)---a member of the class of  causal estimands named the weighted average treatment effects (WATEs). \cite{hirano2003efficient} Unlike the ATE, the ATO bypass the demand on the positivity assumption, but moves the goalpost {which is the target of inference (i.e., the causal estimand)} and applies to a weighted population called the ``equipoise population'', instead of overall population targeted by ATE. \cite{matsouaka2020framework, lawrance2020estimand} The principle of the ATO is simple: rather than giving larger weights to participants whose PS values are at the extremities of the PS spectrum, it gradually and smoothly down-weights their contribution to 0 as their PS approaches either extremities while giving larger weights the closer the PS is to 0.5. 

In fact, the distribution of the weights and the underlying patient population when using ATO is govern by the function $h(x) = e(x)(1-e(x))$, where $e(x)$ is the PS. Since $h(x) = e(x)(1-e(x))$ reaches its maximum at $e(x) = 0.5$ and $h(x)\in [0.16,0.25]$ when $e(x)\in [0.2, 0.8]$, ATO assigns more weights to participants whose PSs are inside the interval $[0.2,0.8]$. {Otherwise}, ATO “gradually and smoothly” down weight the contributions of the participants, the further their PS move away from $[0.2,0.8]$. Thus, targeting more patients in the vicinity of 0.5, {based on their covariates, emulates the traits} of clinical equipoise as in the context of randomized clinical trials. \cite{matsouaka2020framework, zhou2020propensity} 
The new target population is clinically meaningful \cite{thomas2020overlap} and allows a number of practical applications. \cite{li2013weighting,matsouaka2020framework,zhou2020propensity,matsouaka2024overlap} Furthermore, the ATO avoids the constraint of choosing ad-hoc trimming or truncation threshold(s). {The ATO also the advantage of providing better internal validity than ATE, e.g., less sensitive to PS model misspecification, \cite{zhou2020propensity} and thus it has important and often useful practical applications.\cite{matsouaka2020framework,mao2018propensity}}

Issues related to violations of the positivity assumption are also present when we estimate the average treatment effect on the treated (ATT).  ATT reflects ``how would the average outcomes have  differed on treated participants, had they not (counter to the fact) received the treatment?''\cite{greifer2021choosing} In policy-making, ATT helpful in deciding whether a policy currently implemented on some patients should continue or even extended to larger group of participants. To identify ATT from observational data also requires a positivity assumption. {Fortunately, to estimate ATT only requires a weaker positivity assumption, i.e., the PSs for control participants must be smaller than 1.} While there is a plethora of methods for dealing with the lack of positivity when estimating ATE,   there is surprisingly no such systematic effort when estimating  ATT when the positivity assumption is (nearly) violated. 
So far, only Ben-Michael and Keele \cite{ben2023using} compared the performance of OW when estimating the average treatment effect on the overlap (ATO) to that of a balancing weight method (which solves a convex optimization
problem to find weights that aim directly at covariate balance\cite{chattopadhyay2020balancing}), which targets ATT. As indicated, there are some limitations. When using OW, in general, it is only when the treatment effect is constant that the OW can target ATT, which it is too restrictive. In addition, the balancing weight method {they considered} was not superior {to its comparators} when the positivity is poor. 

In this paper, we {explore alternative solution to estimate ATT (or ATT-like) estimand under violations of the positivity assumption.} First, we review commonly-used methods, such as trimming and truncation, when used to estimate ATT. Following the framework of the WATE estimation, we discuss how to apply them to target ATT, since current practices do not have a common standard on trimming and truncation for ATT. Sometimes their applications overlook some basic tenets of elucidating their goalposts, i.e., what underlying population and estimand they should target.  Our main contribution is the proposal of a PS-weight alternative for average causal effect on the treated, called overlap weighted average treatment effect on the treated (OWATT), by leveraging the framework of OW. 

The remainder of this paper is organized as follows. In Section \ref{sec:setup}, we introduce the standard methods for estimating ATT. Then, we review the current practices when dealing with violations (or near violations) of the positivity assumption for ATT. In Section \ref{sec:watt}, we introduce our proposed method, present the corresponding estimator, and study its inference. In Section \ref{sec:sim}, we assess the proposed method in a simulation study, with a wide range of scenarios, and compare it to existing methods. In Section \ref{sec:data}, we illustrate the proposed method to investigate racial disparities in health care expenditures using data from the
Medical Expenditure Panel Survey (MEPS). Finally, in Section \ref{sec:remarks} we conclude the paper with remarks, suggestions, and future directions. 

\section{Definitions and Notations}\label{sec:setup}
\subsection{The average treatment effect on the treated (ATT)}\label{subsec:framework}
Consider a non-randomized study, with a treatment assignment indicator $Z\in\{0,1\}$ that determines whether a participant received the treatment ($Z=1$) or a  control  ($Z=0$). Let $Y$ denote the observed continuous outcome, and  ${X}=(X_1, \ldots, X_p)'$ denote the vector of measured covariates. The observed data $\mathcal{O}=\{ (Z_i, X_i, Y_i): i=1,\dots, N \}$ represent an independently identically distributed (i.i.d.)  sample of $N$ participants from a large population. 

We follow the potential outcome framework of Neyman-Rubin \citep{neyman1923applications, imbens2015causal} and assume that each participant has two (ex-ante fixed, but a {priori} unknown) potential outcomes associated with treatment assignment, denoted by $Y_i(z)$ for $z=0,1$, where  $Y_i = Z_iY_i(1) + (1-Z_i)Y_i(0)$. The potential outcome $Y_i(z)$ represents the outcome a participant $i$ will experience if, possibly contrary to fact, assigned to treatment $Z_i=z_i$. 

Although a participant's causal treatment effect is defined as $Y_i(1) -Y_i(0)$, we cannot observed both $Y_i(1)$ and $Y_i(0)$ simultaneously. Instead, we must rely on a sample of participants to estimate average treatment effects. For this purpose,  we assume the stable-unit treatment value assumption (SUTVA), i.e., there is only one version of the treatment, and the potential outcome $Y_i(z)$ of an individual does not depend on and does not impact another individual's received treatment. \citep{rosenbaum1983central}

The PS defined by a subject-specific probability of receiving the treatment given the subject's covariates, i.e.,  $e_i({x})=P(Z_i=1\mid {X_i=x_i})$, will play a major role in this paper. The PS is unknown for a non-randomized study;  we usually postulate a (parametric) model (e.g., using a generalized linear model (GLM)) to estimate it.  Hereafter, we assume the PS model is correctly specified by a GLM and use the notation $e_i(X_i; \beta)$ where $\beta$ is the vector of regression coefficients for the covariates in the model. The correct model specification means that there exists $\beta^*$ in the space of parameters $\beta$  such that $e_i(X_i; \beta^*)$ is indeed the true PS. The estimated PS is then given by $\widehat e_i(X_i) = e(X_i; \widehat\beta_N)$ where $\widehat\beta_N$ is a consistent estimator for $\beta$.  

Our main estimand of interest, the average treatment effect on the treated (ATT), is defined by $$\tau_{att} = \Ex\{Y(1)-Y(0)\mid Z=1\}=\Ex\{Y\mid Z=1\}-\Ex\{ Y(0)\mid Z=1\}.$$
The first term $\Ex\{Y\mid Z=1\}$ is potentially observable and can be estimated from the data. However, the second term, $\Ex\{Y(0)\mid Z=1\}$---the average outcome among treated participants had they not been treated---is not observable since for any particular treated participant, we cannot observe (i.e. measure) $Y(0)$. The main idea of most causal inference methods in estimating ATT is to leverage measured data from control participants to impute missing the counterfactuals $Y_i(0)$ in treated participants (i.e., as good proxies) to provide a consistent estimate of  $\Ex\{Y(0)\mid Z=1\}$  by making some specific, often untestable, assumptions. 

To estimate ATT using PS weighting methods, there are two important assumptions to be made. 
\begin{enumerate}%[label=(\alph*)]
\item\textsl{Unconfoundness}: $\Ex\{Y(0)\mid  Z=0,X\} = \Ex\{Y(0)\mid  Z=1,X\} $\label{ref:uncounfond}
\item \textsl{Positivity}: (a) $P(Z=1)>0$, and (b) $P(e(X)<1)=1$ on control participants.\label{ref:positivity}
\end{enumerate}
\textbf{Remarks. } Although the Assumption \ref{ref:uncounfond} is weaker than the uncounfoundness assumption often used in the literature, it however suffices (along with Assumption \ref{ref:positivity})  to identify ATT. Assumption \ref{ref:positivity}(a), $P(Z=1)>0$, is trivial as we need a fraction of the population to receive the treatment to estimate ATT. The assumption \ref{ref:positivity}(b),  $e(X)<1$ with probability 1 on control participants, implies that the support $Supp(X\mid Z=0)$ of covariates for the participants under control contains the support  $Supp(X\mid Z=1)$  of treated participants {(see Abadie \citep{abadie2005semiparametric} and Heckman et al. \citep{heckman1998matching} for important insights on these conditions).} Note that the above Assumption \ref{ref:positivity}(b) is an identification, which is not needed in other estimation methods of ATT, such as matching where the focus is on the common support $Supp(X\mid Z=1)\cap Supp(X\mid Z=0)$. For matching, the impetus is to find matches that are fairly similar; thus, we emphasize a good overlap of the PS distributions between the treatment groups.\cite{stuart2010matching}

Under these two assumptions, it can be shown that (see Appendix \ref{subapx:equi-form-att})
\allowdisplaybreaks \begin{align}\label{eq:att}
    \tau_{att} = \frac{\Ex(ZY)}{\Ex(Z)}  - \frac{\Ex\left\{w_0(X)(1-Z)Y\right\}}{\Ex\{w_0(X)(1-Z)\}}, 
\end{align}
where $w_0(X)={e(X)}\{1-e(X)\}^{-1}$. These conventional ATT weights $w_0(X)$
generate a pseudo-population among the control participants by re-weighting their contributions, \cite{hirano2003efficient, li2018balancing} which helps the covariate distribution of these control group participants look as similar as possible to that of the treatment participants. Thus, by Assumption \ref{ref:uncounfond} (unconfoundness), the (weighted) observed outcomes of the control participants can serve as counterfactual alternatives of $Y_i(0)$ for treated participants $i$ to finally estimate $\tau_{att}.$   In a sense, the weights explicitly indicate the contribution of each control participant in approximating such counterfactuals, based on their covariate distributions. Thus, using the group of the treated participants as reference,  control participants are either up-weighted or down-weighted, depending on whether their PSs are similar to those of treated participants: the higher the PS, the higher the weight. %who are less likely to be treated (i.e., with small   are down-weighted, whereas those who are more likely to be treated (had they had a chance) based on their covariates $X_i$ (i.e., with high values of $e(X_i)$)  are upweighted.  %More specifically, define the weights

An estimator of  $\tau_{att}$ is given by
\allowdisplaybreaks \begin{align}\label{eq:ipw-att}
    \widehat\tau_{att} & = \frac{\displaystyle\sum_{i=1}^N Z_iY_i}{\displaystyle\sum_{i=1}^N Z_i} - \frac{\displaystyle\sum_{i=1}^N (1-Z_i)\widehat w_0(X_i)Y_i}{\displaystyle\sum_{i=1}^N (1-Z_i)\widehat w_0(X_i)},
\end{align}
where  $\widehat w_0(X_i)={\widehat e(X_i)}\{{1-\widehat e(X_i)}\}^{-1}$.  
Such a normalized estimator %(i.e.,  whose weights of the outcome $Y$ sum up to 1 within each treatment group)
{(i.e., where the weights within each treatment group  sum up to 1)}
such as  \eqref{eq:ipw-att} is often preferred in finite sample as it is more stable and more efficient. \cite{busso2009finite,matsouaka2020framework, matsouaka2023variance} 

When the PS model is correctly specified, the above estimator is consistent for ATT and has an asymptotic normal distribution. It can be seen that the positivity assumption is extremely important since large PSs (close to 1) for the control participants will result in extreme values of $w_0(x)$. Under finite-sample, extreme weights may incur an inefficient estimate. In addition, it has been shown that the estimator \eqref{eq:ipw-att} is sensitive to extreme weights and model misspecifications. \citep{matsouaka2024overlap, matsouaka2023variance}

\subsection{Addressing extreme weights through trimming or truncation}\label{subsec:trimming}

Two widely-used strategies for dealing with the lack of positivity (i.e., when extreme weights exist) are PS trimming and truncation. {ATT} trimming drops all control participants whose estimated PSs are outside of the interval $[0,1-\alpha]$ from the study sample, with  $\alpha\in(0,0.5)$  a user-specified threshold. Because trimming is not smooth, one of its key drawbacks is that the use of bootstrap to estimate the variance is not always valid, as the condition needed to adequately apply the bootstrap technique does not hold.\cite{yang2018asymptotic} %Sometimes the confidence intervals can be too large; sometimes too small.
Instead of abruptly discarding control participants whose PS is above $1-\alpha$, Yang and Ding \cite{yang2018asymptotic} propose to reduce their weights gradually and smoothly toward 0 (but still approximate the above trimmed sample) and derive an estimator that has better asymptotic properties. With truncation, instead of reducing the weights $w_0(x)={e(x)}\{{1-e(x)}\}^{-1}$  of control participants whose PSs fall outside of the interval  $[0,1-\alpha]$ to 0, they are replaced by a constant value $w_0(x)=({1-\alpha})\alpha^{-1}$. 

There have been two approaches for {ATT} trimming (resp. {ATT} truncation) participants in the literature. One is to trim (resp. truncate) both the treated and control participants, the other is to only trim (resp. truncate) the control group. Heckman et al.\cite{heckman1998matching, heckman1998characterizing} and Smith and Todd\cite{smith2005does} excluded all observations in both the treated and control groups with estimated PS of {1 or near 1}. Yang and Ding argued to trim both the treated and control groups, because the control units with estimated PS close to 1 lead to large weights while the treated units with large estimated PS have few counterparts in the control group to make inference on.\cite{yang2018asymptotic} Austin excluded any participant with an estimated PS outside the range of [$\alpha$, $1-\alpha$].\cite{austin2022bootstrap} 

In contrast, Lee et al. set all weights in the control group above a certain threshold equal to the threshold. \cite{lee2011weight} Dehejia and Wahba discarded the units in the control group whose estimated PS was smaller than the smallest estimated PS in the treated group. \cite{dehejia1999causal} Crump et al. followed Dehejia and Wahba by dropping participants in the control group only, but those with estimated PS greater than a certain threshold.\cite{crump2006moving}

In this paper, we trim the  control participants whose estimated PS falls outside the range $[0,1-\alpha]$, but leave the treated group intact. In addition, after trimming, we considered two specific strategies: (1) not to re-estimate the PS (for instance, as adopted by Sturmer et al. \cite{sturmer2010treatment}) and (2) to re-run the PS model on the trimmed sample and re-estimate PSs, following the recommendation from Li et al.\cite{li2019addressing} %In later simulation results (see Section \ref{subsec:results}), re-estimating the PS results in smaller biases of the ATT trimming estimator when the PS model is misspecified, but it indeed introduces more biases when the PS model is correctly specified. 

%%%%%%%%%%%%%%%%%%%%%%%%%%%%%%%%%%%%%%%%%%%%%%%%%%%%%%%%%%%%%%%%%%%%%%%%%%%%%%%%%%%%%%%%%%%%%%%%%%%%%%%%%%%
\section{Weighted Average Treatment Effect on the Treated}\label{sec:watt}

\subsection{{ATT} trimming and truncation as scaling techniques}\label{subsec:trim_watt}

The above {ATT} trimming and {ATT} truncation estimators (in Section \ref{subsec:trimming}) proposed to deal with issues related to violations of the positivity Assumption \ref{ref:positivity} are all scaled version of the ATT, in the sense that the weights $w_0(X)$ for control participants are scaled by a multiplying factor $h(X)$ (a function of the PS $e(X)$) to circumvent such violations. Indeed, these methods change the weights on the control in the estimand \eqref{eq:att}, going from $w_0(X)$ to $w_0(X)h(X)$. In this regard, they estimate the treatment effect on the treated over a subset $\mc H$ of the support of $X$. 

% The trimming estimators  share the same  $\mc H=\{X: 0\leq e(X)<1-\alpha\}$. Their functions $h_0$ are given by $h(X)=%\mathcal{I}
% \bd 1(0\leq e(X)<1-\alpha)$ for the standard trimming and by   $h(X)=\Phi_{\varepsilon}(1-e(X)-\alpha)$,  for the smooth trimming proposed by Yang and Ding, where $\Phi_{\varepsilon}$ is the  is the normal cumulative distribution with mean zero and variance $\varepsilon^2,$ for some $\varepsilon>0$.  Note in their paper, Yang and Ding also assign a  weight equal to $h(X)=\Phi_{\varepsilon}(1-e(X)-\alpha)$ to treated participants, which is unnecessary since $E[Y(1)\mid Z=1]=E[ZY]$ , i.e., each treated participant has a weight of 1.

%For trimming estimators defined in Section \ref{subsec:trimming} have the following illustrations in the WATT class. 
First, for the standard (non-smooth) {ATT} trimming with threshold $\alpha\in(0,0.5)$, we can either use (i) $\mc H=\{(X,Z):0\leq Ze(X)\leq 1, 0\leq (1-Z)e(X)\leq 1-\alpha\}$ and $h(X)=1$, or (ii)  $\mc H=\{X:0\leq e(X)\leq 1\}$ and ${h(X)=\bd 1\{0\leq e(X)\leq 1-\alpha\}}$ to define its target estimand. Illustration (i) shows that trimming is applying ATT to the subpopulation defined by $\mc H$ , whereas (ii) indicates that trimming can also be viewed as using a non-smooth function to weight on the control group. 

For smooth {ATT} trimming,\cite{yang2018asymptotic} with a threshold $\alpha$, we have $\mc H=\{X:0\leq e(X)\leq 1\}$ and $h(X)=\Phi_{\varepsilon}(1-e(X)-\alpha)$ where, for some $\varepsilon>0$, $\Phi_{\varepsilon}$ is the cumulative distribution function (CDF) of a normal distribution, with mean zero and variance $\varepsilon^2.$ Whenever $\varepsilon\to 0$, $h(X)\to {\bd 1\{0\leq e(X)\leq 1-\alpha\}}$, i.e., for some small $\varepsilon$, this method approximates standard {ATT} trimming (based on the above illustration (ii)), but using a smooth weight function (everywhere) on the control. In other words, smooth {ATT} trimming cannot be illustrated using (i) above, because it does not really trim samples with $e(X)>1-\alpha$, but applies on them a weight that is {very} close to 0. 

Finally, for {ATT} truncation,  with a threshold $\alpha,$ we have  $\mc H=\{X: 0\leq e(X)\leq 1\}$ and $h(X)=\bd 1(0\leq e(X)<1-\alpha) + (1-\alpha)\alpha^{-1}{w_0(X)}^{-1}\bd 1\{e(X)\geq 1-\alpha\} $.

Overall, following the weighted average treatment effects (WATE) framework,\cite{li2018balancing,matsouaka2020framework} all the above estimands can be wrapped up into a class of generalized versions of ATT, called \textit{weighted average treatment effect on the treated (WATT)}, and defined by
\allowdisplaybreaks \begin{align}\label{eq:watt}
    \tau_{watt}^h(\mc H)
    & = \frac{\Ex(ZY)}{\Ex(Z)} - \frac{\Ex\left\{\omega_{0h}(X)(1-Z)Y\right\}}{\Ex\{\omega_{0h}(X)(1-Z)\}},~~\text{with}~~ \omega_{0h}(x)=w_0(x)h(x).
\end{align}
The function $h(x)$, hereafter the \textit{tilting function}, specifies the weighted target population $\mc H$ for control participants, which is  a subset $\mc H$ of the support $Supp(X\mid Z=0)$ of the covariates $X$. By abuse of language, the estimand $\tau_{watt}^h(\mc H)$ can also be referred to as the treatment effect on the treated over the subset $\mc H$, whenever the context allows.

To identify WATT from the observed data, we propose the following simple weighted estimators:
\allowdisplaybreaks \begin{align}\label{eq:wt-watt}
    \widehat\tau_{watt}^h(\mc H)& = \frac{\displaystyle\sum_{i=1}^N Z_iY_i}{\displaystyle\sum_{i=1}^N Z_i} - \frac{\displaystyle\sum_{i=1}^N (1-Z_i)\widehat \omega_{0h}(X_i)Y_i}{\displaystyle\sum_{i=1}^N (1-Z_i)\widehat\omega_{0h}(X_i)}
\end{align}
where $\widehat\omega_{0h}(X_i)=\widehat w_0(X_i)\widehat h(X_i)$ and $\widehat h(X_i)$ is calculated by plugging $\widehat e(X_i)$ into the function $h(X_i)$. 
Evidently, when $h(X)\propto 1$,
%and $\mc H=\{X:0\leq e(X)\leq 1\}$, 
$\tau_{watt}^h(\mc H)$ and $\widehat\tau_{watt}^h(\mc H)$ simplify to $\tau_{att}$ and $\widehat \tau_{att}$, respectively. Furthermore, by classic asymptotic theory, we can show that when the PS model is correctly specified, $\widehat\tau_{watt}^h(\mc H)$ is a consistent estimator of  $\tau_{watt}^h(\mc H)$. 

%{
% We need to explain why when we do trimming, truncation, and smooth trimming, we also change the denominator to be $\Ex\{\omega_{0h}(X)(1-Z)\}$, but not $\Ex(Z)$. 

% My crude explanation of this now is that: our definition of trimming and truncation in this way mimics the WATE analogy for trimming when using indicator for its tilting function. In addition, the weighted effect on the control participants also makes the estimand ``closer'' to ATT, otherwise what it is targeting for is also harder to interpret, and incurs substantial bias to ATT, or affects our knowledge about the true effect. 

%This is just some initial thought using crude language and may not be comprehensive, and we need to think about more aspects and how to write them in the paper finally. }

% To contrast with the ATT estimand \eqref{eq:att}, we will refer to the estimand $\tau_{att}(\mc H)$ as the weighted treatment effect on the treated (WATT). 
By using either {ATT} trimming or truncation, we implicitly make the decision to change the target of interest, \cite{crump2006moving} as shown by the above formulas \eqref{eq:watt} and \eqref{eq:wt-watt}.
{ATT} trimming (resp. truncation) excludes (resp. stabilizes) the weights of control participants whose PS are near or equal to 1. Unfortunately, these methods are ad-hoc and thus subjective; they  depend on the choice of the user-specified threshold  $\alpha$, which can have a substantial impact on the estimation results. Furthermore, for the smooth {ATT} trimming,\cite{yang2018asymptotic}  we also need to select an appropriate value for $\varepsilon$. 

Although Yang and Ding\cite{yang2018asymptotic} suggested using a small $\varepsilon$, such as   $\varepsilon = 10^{-4}$ or  $10^{-5}$, %these values or smaller do not make much of difference compared to the standard trimming in terms of the point estimate.Its only competitive advantage it the possibility to. 
there is actually a bias-variance trade-off that needs to be made. While choosing smaller $\varepsilon$ helps achieve a better estimate of $\tau_{watt}^h(\mc H)$, when the PSs are correctly estimated, it also leads to weights that are closer or similar to those from the standard {ATT} trimming and increases the variance of $\widehat \tau_{watt}^h(\mc H)$. Thus, it is often recommended to run  a sensitivity analysis by varying  $\varepsilon$ over a grid of possible values. 

For both {ATT} trimming and truncation, there is not a general consensus on which value of $\alpha$ (and  $\varepsilon$) achieves the optimal bias-variance trade-off. \citep{crump2009dealing,yang2018asymptotic} Similar concerns have been voiced in the estimation of the average treatment effect.\cite{zhou2020propensity,matsouaka2020framework} Therefore, given the broad discretion to subjectively choose a threshold $\alpha$ (and the smoothing parameter  $\varepsilon$), there is a higher risk to cherry-pick and report results that suit us,  hence falling into the trap we wanted to avoid by using a PS method in the first place.\citep{rubin2007design,rubin2008objective}

What should we do then? We propose the use of a method that is robust to and independent of such user-specified threshold and parameters.

\subsection{Overlap weight average treatment effect on the treated (OWATT)}\label{subsec:owatt}
We consider $\mc H=\{X: 0\leq e(X)\leq 1\}$ and $h(X)= e(X)\{1-e(X)\}$, i.e., the tilting function of the overlap weight estimator. \cite{li2018balancing} We refer to   $\tau_{watt}^h(\mc H)$ as the overlap weight average treatment effect on the treated (OWATT). 
It circumvents the need to specify a threshold $\alpha$ (and $\varepsilon$), but still provides smooth weighting (without extreme weights or the need for {ATT} trimming), and has better asymptotic properties. \citep{hirano2003efficient, li2018balancing}

The function $h(X)=e(X)\{1-e(X)\}$ targets the population of participants for whom there is a clinical equipoise, i.e., those who have high probabilities of receiving either treatment option. Since $h(X)$ reaches its maximum value at $e(X)=0.5$, it weighs the most participants at ``equipoise''\citep{matsouaka2020framework,matsouaka2024overlap} (mimicking 1:1 allocation in a randomized trial). The weights $\omega_{0h}(X)$ simplifies to $e(X)^2$ and do not involve $(1-e(X))^{-1}$, which circumvent {the issue of large weights altogether.} 

As function of $e(X)$,  the weights $\omega_{0h}(X)=e(X)^2$ are monotone increasing  throughout the whole PS domain $\{e(X): 0\leq e(X)\leq 1\}$. They reach their highest value at $e(X)=1$, which contrast drastically with the behaviors of {ATT} truncation and {ATT} trimming weights, in the region $\mc R=\{e(X): e(X)\geq 1-\alpha\}$. For {ATT} truncation $\omega_{0h}(X)=(1-\alpha)\alpha^{-1}$ is constant on $\mc R$, regardless of the individual PSs, while the {ATT} trimming weights are either 0 (standard trimming) or slowly decrease to 0 (smooth {ATT} trimming). Furthermore, when  $e(X)$ approaches 1, oddly enough, the weights for smooth {ATT} trimming go back up to infinity, defeating the purpose of why {ATT} trimming was used in the first place (see Figure \ref{fig:tilt_wts} below).

Why are these properties of $\omega_{0h}(X)=e(X)^2$ important? The goal of any PS weighting to estimate ATT-like estimands is to judiciously use the contributions of control participants and, ultimately, impute missing counterfactuals outcomes $Y(0)$ for treated participants.
Therefore, the rationale behind PS weighting when estimating ATT (or ATT-like estimand) is to use the group of treated participants as the reference group from which the controls cases are standardized to, based on their PSs: control participants whose covariates are most similar to the reference group are up-weighted while those with dissimilar covariates are down-weighted. Hence, of all the weights we consider in this paper, only the overlap weights follow this rationale and continue to up-weight the contribution of control participants whose characteristics, encapsulated by the PSs, are similar to those of treated participants. When $e(X)=0,$ all the $\omega_{0h}(X)$'s are equal to 0, including $\omega_{0h}(X)=e(X)^2=0$ {for OWATT}. However, when  $e(X)=1$, $\omega_{0h}(X)$ is still equal to 0 for standard {ATT} trimming and remains constant (and equal to $(1-\alpha)/\alpha$) for {ATT} truncation, but $\omega_{0h}(X)$ goes to infinity when we do not trim or when we use smooth {ATT} trimming ATT estimators, for some values of $\varepsilon$.  

\begin{figure}
    \centering
    \includegraphics[trim=10 10 10 10, clip, width=0.8\textwidth]{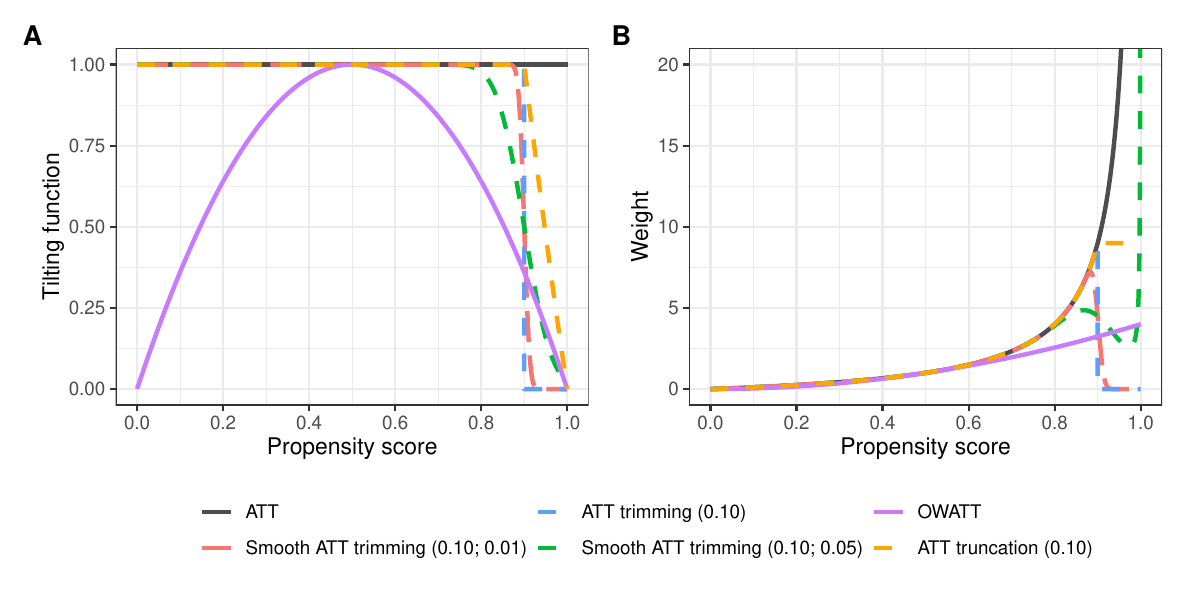}
    
    \begin{tablenotes}\footnotesize
     \item  Panel A: tilting functions  $h(x)$; Panel B: PS weights $\omega_{0h}(x) = \dfrac{e(x)h(x)}{1-e(x)}$.  
     \item For trimming or truncation, $\alpha=0.1$; also for smooth trimming $\varepsilon=0.01$ (red line) and $0.05$ (green line).
     \item For OWATT, we use $h(x) = 4e(x)\{1-e(x)\}$  (purple line) for illustration and comparison purposes. 
    \caption{Tilting functions and propensity score weights of control participants for different estimands}
    \label{fig:tilt_wts}
\end{tablenotes}\end{figure}

As noted from Figure \ref{fig:tilt_wts}, the smooth ATT trimming with $\alpha = 0.1$ and $\varepsilon=0.05$ has weights that decrease in the region $\{e(X): e(X)\in [1-\alpha, 1)\}$, but go sharply back to infinity as soon as  $e(x)$ is {fairly close or equal to} 1. The reason being that when $e(x)$ is close or equal to 1, $\Phi_{\varepsilon}(1-e(X)-\alpha) = \Phi_{0.05}(0.90-e(x))$ does not equal or approximate  0. Thus, the weights are still dominated by ${e(x)}\{1-e(x)\}^{-1}$ which remains large when $e(x)\approx 1$. Thus, oddly enough, the smooth trimming method sometimes may still suffer from the extreme weights when the duo $(\alpha,\varepsilon)$ is not selected appropriately. However, this {behavior is not observed for instance} when $\varepsilon=0.01$.

\subsection{Inference}\label{subsec:inference}

Using asymptotic theory, we can show that when the PS model is correctly specified, $\widehat\tau_{watt}(\mc H)$ is consistent to $\tau_{watt}^h(\mc H)$.  Furthermore, we have the following asymptotic result: 
\begin{theorem}\label{th:watt}
    Under some regularity conditions, the estimator $\widehat\tau_{watt}^h(\mc H)$ is regular and asymptotic linear (RAL), if the propensity score is specified by some generalized linear model $e(X)=e(X;\beta)$ with $\beta^*$ the true value of $\beta$. Furthermore, $$\sqrt{N}(\widehat\tau_{watt}^h(\mc H)-\tau_{watt}^h(\mc H))\to_d\text{N}(0,\sigma^2 + b_1'\mc I(\beta^*)^{-1}b_1 - b_2'\mc I(\beta^*)^{-1}b_2),$$ where  $\displaystyle \sigma^2  = \sum_{z=0}^1\Ex\left\{\eta_z(X)\{\mu\{z,e(X)\}^2 + \sigma^2\{z,e(X)\} + \sigma^2(z,X)\}\right\}$ with  
  \allowdisplaybreaks   \begin{align*}
          \eta_1(X)  &= \frac{e(X)}{\Ex\{e(X)\}^2},~\eta_0(X)=\frac{ \omega_{0h}(X)^2\{1-e(X)\}}{\Ex\{e(X)h(X)\}^2},~
         \mu\{z, e(X)\}  = \Ex\{Y\mid e(X), Z=z\},\\
         \sigma^2\{z, e(X)\} &= \var\{Y\mid e(X), Z=z\},\quad\text{and}\quad\sigma^2(z,X)  = \var\{Y\mid X, Z=z\}, \quad\text{ for }z=0,1,
    \end{align*}
    $\mc I(\beta^*)$ is the Fisher's information matrix of $\beta$, and $b_1$, $b_2$ are specified in Appendix \ref{subapx:th1-pf}. 
\end{theorem}
\textit{Proof. } See Appendix \ref{subapx:th1-pf}.

It ensues from Theorem \ref{th:watt} the following remarks:
\begin{enumerate}
    \item The use of the standard bootstrap for variance estimation is valid, because $\widehat\tau_{watt}(\mc H)$ is asymptotic linear. \cite{shao2012jackknife} The empirical performance of bootstrap variance estimation is examined in the simulation study of Section \ref{sec:sim}. 
    \item The above asymptotic variance of $\widehat\tau_{watt}^h(\mc H)$ reveals that the uncertainty to estimate $\tau_{watt}^h(\mc H)$ comes from multiple sources of variability, including the uncertainty to estimate $\beta$ and the variability in observed outcome. The choice of $h(x)$ also affects the variance  of $\widehat\tau_{watt}^h(\mc H)$. For example, when $h(x)=1$, $\omega_{0h}(x) =w_0(x) =e(x)/\{1-e(x)\}$ and $\widehat\tau_{watt}^h(\mc H)$ estimates ATT. If some $e(x)$'s in control group are close to 1, $\eta_0(x)\propto e(x)^2/\{1-e(x)\}$ will be large, thus will also incur large variance. When choosing the overlap function $h(x)=e(x)\{1-e(x)\}$,  $\eta_0(x)\propto e(x)^4\{1-e(x)\}$, which is then bounded and less variable. 
\item  We have shown (see Appendix \ref{subapx:asybeh-miss}) that, when the PS model is possibly misspecified and equal to $\widetilde e(X)$, the asymptotic bias of  $\widehat\tau_{watt}^h(\mc H)$ is 
\allowdisplaybreaks 
\begin{align*}
    \text{ABias}(\widehat\tau_{watt}^h(\mc H)) & = \frac{\Ex\{\omega_{0h}(X)\{1-e(X)\}m_0(X)\}}{\Ex\{\omega_{0h}(X)\{1-e(X)\}\}} - \frac{\Ex\{\widetilde\omega_{0h}(X)\{1-e(X)\}m_0(X)\}}{\Ex\{\widetilde\omega_{0h}(X)\{1-e(X)\}\}}, 
\end{align*}
where $m_0(X) = \Ex\{Y(0)\mid X\}$, and $\widetilde \omega_{0h}(X)$ denotes the limit of $\widehat \omega_{0h}(X)$ using misspecified PS model $\widetilde e(X)$ in lieu of the true PS $e(X)$. 
Clearly, whether  we choose $h(x)=1$ vs. $h(x)=e(x)\{1-e(x)\}$, the asymptotic bias is, respectively,
\allowdisplaybreaks \begin{align*}
    \frac{\Ex\{e(X)m_0(X)\}}{\Ex\{e(X)\}} - 
    \frac{\Ex\left\{\dfrac{\widetilde e(X)}{1-\widetilde e(X)}\{1-e(X)\}m_0(X)\right\}}{\Ex\left\{\dfrac{\widetilde e(X)}{1-\widetilde e(X)}\{1-e(X)\}\right\}}\\
    \text{ or }\quad
    \frac{\Ex\{e(X)^2\{1-e(X)\}m_0(X)\}}{\Ex\{e(X)^2\{1-e(X)\}\}} - 
    \dfrac{\Ex\left\{\widetilde e(X)^2\{1-e(X)\}m_0(X)\right\}}{\Ex\left\{\widetilde e(X)^2\{1-e(X)\}\right\}},
\end{align*}
when we use $\widetilde e(X)$ in lieu of the true PS $e(X)$.  This shows that the latter bias is less sensitive and less extreme to large values of  $\widetilde e(x)$, i.e., when $\widetilde e(x)\to 1$. 
\end{enumerate}
We provide more details on the asymptotic bias when the PS model is (possibly) misspecified in the Appendix \ref{subapx:asybeh-miss}. In the simulation study in Section \ref{sec:sim}, we also compare the performance of the  OWATT estimator against the trimming and truncation estimators, when the PS model is misspecified.

\section{Simulation}\label{sec:sim}

Based on the chosen data generating processes (DGPs) in Section \ref{sec:DGP}, we generated two sets of data. First, we simulated 10 independent superpopulation of $10^6$ individuals (under the different scenarios) to determine the true values of the estimands under heterogeneous treatment effect. {For example, for ATT, we independently generated $10^6$ pairs of potential outcomes $(Y(1), Y(0))$ under the chosen DGPs, and then calculate $\displaystyle\sum_{i=1}^{10^6}Z_i\{Y_i(1)-Y_i(0)\}\bigg/\displaystyle\sum_{i=1}^{10^6}Z_i$ as the true value of $\tau_{att} = \Ex\{Y(1)-Y(0)\mid Z=1\}$. We repeat this process independently 10 times and average the 10 calculated ``true values''  to get the final true value of ATT. By averaging over the  10 independent super-population data, the uncertainty is negligible. Similar calculation applies to other estimands. }

Then, to assess the finite-sample performance of the different estimators, we simulated $M=1000$ data sets of size $N$ and allowed the overlap of the distribution of the PS to vary, as specified below. Within each data set, we estimated the above estimands. For ATT trimming (standard and smooth), we trimmed observations with the PS that fall outside of the interval {$[0, 1-\alpha]$}, with $\alpha = 0.05, 0.1$, and $0.15$, respectively. We summarized the data in tables and figures and interpreted them accordingly, following the measures and evaluation criteria laid out in Section \ref{subsec:compare}. 

{\color{black}We provide R functions implementing the point estimators and bootstrap variance for inference of all methods considered in this paper, with an illustrative example, 
available at \url{https://github.com/yiliu1998/ATTweights}. }

\subsection{Data generating process}\label{sec:DGP}
{\color{black} We considered two different DGPs, one similar to the DGP used by Li and Li\cite{li2021propensity} (DGP I) and the other following the classic DGP of Kang and Schafer\cite{kang2007demystifying} (DGP II). } 

\subsubsection{DGP I:}

{\color{black} We chose $N=1000$ and first generated the  covariate vector $X = (\bd {1},X_1,...,X_7)'$, such that }  $X_4\sim \text{Bern}(0.5)$,  $X_3\sim \text{Bern}(0.4+0.2X_4)$, $(X_1,X_2)\sim \text{BN}(\mu,\Sigma)$, a bivariate normal with $
    \mu = \left(-0.25X_3+X_4+X_3X_4, X_3-0.25X_4+X_3X_4\right)$  and  $
    \Sigma = X_3\begin{pmatrix}
    1&0.5\\0.5&1
    \end{pmatrix} + 
    (1-X_3)\begin{pmatrix}
    2&0.25\\0.25&2
    \end{pmatrix}
    $, while $X_5=X_1^2$, $X_6=X_1X_2,$ and $X_7=X_2^2.$ 
    
Next, we considered $\alpha =(\alpha_0, \alpha_1,\dots,\alpha_7)$ where $\alpha_1=...=\alpha_4=-0.4\gamma$ and $\alpha_6=\alpha_7=-\alpha_5 = 0.1\gamma$ to generate the treatment assignment $Z\sim \text{Bern}(e(X))$ based on the PS model $e(X) = \{1+\exp(-X'\alpha)\}^{-1}.$ By varying $\gamma$, we generated three distinct scenarios for the overlap of the distributions of the PS between the two treatment groups,   labeled, respectively, good, moderate, and poor overlap. Moreover, we judiciously selected the intercept $\alpha_0$  to have roughly the same numbers of participants between the two treatment groups. Hence, we chose $(\alpha_0,\gamma) = (0.5,0.5)$  \textcolor{black}{(Good overlap)},  $(\alpha_0,\gamma) = (1.2,1.5)$  \textcolor{black}{(Moderate overlap)}, and $(\alpha_0,\gamma) = (2.1,2.5$) \textcolor{black}{(Poor overlap)}. The above PS model was considered misspecified when we exclude the higher order terms $X_5=X_1^2$, $X_6=X_1X_2,$ and $X_7=X_2^2.$ %\textcolor{black}{In the following, we refer $X_1$--$X_4$ as the main covariate terms, since $X_5$--$X_7$ are their higher order terms. }

%\textcolor{red}{The entire paragraph below should be revised. It must be presented in a way that someone who want to re-run these simulations can easily know what to do. So far, it is just very confusing. It is hard to know what models we're talking about, what are the coefficients, what are the true outcome models, and what models specifically produce the four adjusted $R^2$ that are listed. I was unable to figure this out!}

We also generated the observed outcome $Y=ZY(1)+(1-Z)Y(0)$ using the potential outcomes such that $Y(0) = 0.5+X_1+0.6X_2+2.2X_3-1.2X_4+X_5+2X_6+X_7 + \varepsilon \text{ and } Y(1) = Y(0)+\delta(X)$ where $\varepsilon\sim \text{N}(0,2^2)$ and $\delta(X) = 4+3X_5+6X_6+3X_7+X_1X_3$. 

\textcolor{black}{The choice of the outcome regression model $Y(0)$ and its parameters resulted in an adjusted $R^2=0.97$ when we fit the full $Y(0)$ model and an adjusted $R^2= 0.58$ if we only include the main terms $X_1$--$X_4$. These are higher than what we often see in practice (i.e., $0.1\leq R^2\leq 0.3$). Therefore, we additionally considered a case where the adjusted $R^2$ is smaller by increasing the variance of the random error term $\varepsilon$ by $12^2$. This lead to the adjusted $R^2=0.48$ for the full model and  adjusted $R^2=0.27$ for the restricted model where we only include $X_1$--$X_4$. The smaller adjusted $R^2$ case corresponds to the same PS model under the poor overlap case aforementioned.} Nevertheless, \textcolor{black}{our choice of the higher $R^2$ case} was made purposefully to help us focus on the ability of the methods considered, enhance the contrasts and similarities of their performance, and capture their significant differences when the PS or outcome model is misspecified.

{\color{black}
\subsubsection{DGP II:}

This DGP follows closely that of Kang and Schafer,\cite{kang2007demystifying}  a well-known paper discussing the model misspecifications. We first generated the vector of covariate $V=(V_1, \dots, V_4)'$, where $V_i\sim\text{N}(0,1)$ i.i.d., $i=1,\dots,4$. The true PS is generated by $e(V) = \{1+\exp(-V'\alpha)\}^{-1}$, and $Z\sim \text{Bern} (e(V))$, where $\alpha=(-1,0.5,-0.25,-0.1)$. 

In addition, we generated the two potential outcomes as $Y(0)=200+27.4V_1+13.7V_2+13.7V_3+13.7V_4+\varepsilon$, $Y(1)=Y(0)+20$, and $\varepsilon\sim\text{N}(0,1)$ is an independent random error term. 
% \begin{itemize}[noitemsep]
%     \item Four covariates: $V_1, V_2, V_3,$ and $V_4$ such that $ V_i\sim\text{N}(0,1)$.
%     \item The PS model and treatment assignment: $e(V) = \{1+\exp(-V'\beta)\}^{-1}$, and $Z\sim \text{Bern} (e(X)$, where $\beta=(-1,0.5,-0.25,-0.1)'$ and $V =(V_1, V_2, V_3, V_4)'$. 
%     \item Two potential outcomes: $Y(0)=200+27.4V_1+13.7V_2+13.7V_3+13.7V_4+\varepsilon$, $Y(1)=Y(0)+20$, and $\varepsilon\sim\text{N}(0,1)$ is an independent random error term. 
%     \item Four additional non-linear covariates, functions of the covariate vector $V$: $X_1=\exp(V_1/2)$, $X_2=V_2/(1+\exp(V_1))+10$, $X_3=(V_1V_3/25+0.6)^3$, and $X_4=(V_2+V_4+20)^2$. 
% \end{itemize}

Moreover, we generated another covariate vector $X=(X_1,\dots,X_4)'$ which is a function of the covariate vector $V$ as follows. $X_1=\exp(V_1/2)$, $X_2=V_2/(1+\exp(V_1))+10$, $X_3=(V_1V_3/25+0.6)^3$, and $X_4=(V_2+V_4+20)^2$. The PS model $e(V)$ above is the correctly specified model. The model that uses of $X =(X_1, X_2, X_3, X_4)'$ in lieu of $V$ is considered as a misspecified PS model. 
Finally, we considered two specific sample sizes $N\in\{200, 1000\}$, with the number of replicates $M=1000$ to match the choices in Kang and Schafer.\cite{kang2007demystifying}
}

% %\textcolor{red}{Beyond this point, I don't understand what we are trying to do or prove.
% 1. Why didn't we keep the same intercept as in the original Paper by Kang and Schafer?
% 2. Where is the treatment variable in the outcome model? If there isn't, how can you calculate Y(1) and Y(0)?
% 3. What the modification is supposed to do, add, or improve?
% 4. Why do we care about the ATE when the paper is on the ATT estimation?
% 5. What "...matching the potential outcomes framework" mean and why is it important here?
% 6. What is the "...spirit of their DGP"? Why do we need to preserve it here?
% 7. What the point of this paragraph below? Are we trying to hide anything?
%}

\subsection{Performance criteria}\label{subsec:compare}
To evaluate and compare the performance of the estimators considered (and their sensitivity to model misspecifications), we used 3 specific measures: {the absolute relative percent bias (ARBias\%) $=  100\% *\left\vert  \text{mean of }\left(\widehat\tau_{watt}^h-\tau_{watt}^h\right)/\tau_{watt}^h\right\vert$, the 
root mean square error (RMSE) $= \left\{\text{mean of } \left(\widehat\tau_{watt}^h-\tau_{watt}^h\right)^2\right\}^\frac{1}{2}$, and the 95\% coverage probability (CP), i.e., and the proportion of times  $\tau_{watt}^h$ falls inside of its estimated confidence interval, over the $1000$ Monte Carlo replicates.} 

We used $R=200$ standard bootstrap samples, in each simulated data replicate, and calculated the empirical variance of the 200 bootstrap point estimates as the variance estimate. We use ARBias to meaningfully compare different estimators, via tables or graphs, since the underlying causal estimands are expected to be different across methods; the smaller ARBias\%,  the measure, the better.  Also, since we used $M=1000$ simulated data, a 95\% CP is considered significantly different from the nominal 95\% level if it is outside of the interval $[0.937, 0.963].$  

In the Appendix \ref{apx:simu}, we also report 
the relative root mean square error (RRMSE) 
$= \left\{\text{mean of }\left[({\widehat\tau_{watt}^h-\tau_{watt}^h})/{\tau_{watt}^h}\right]^2\right\}^\frac{1}{2}$ and the relative efficiency (RE), i.e., the mean of the ratios of the empirical variance over the bootstrap variance estimate of $\widehat\tau_{watt}^h$ over the $1000$ Monte Carlo replicates.

\begin{figure}[H]
    \centering
    \includegraphics[trim=10 10 5 5, clip, width=0.95\textwidth]{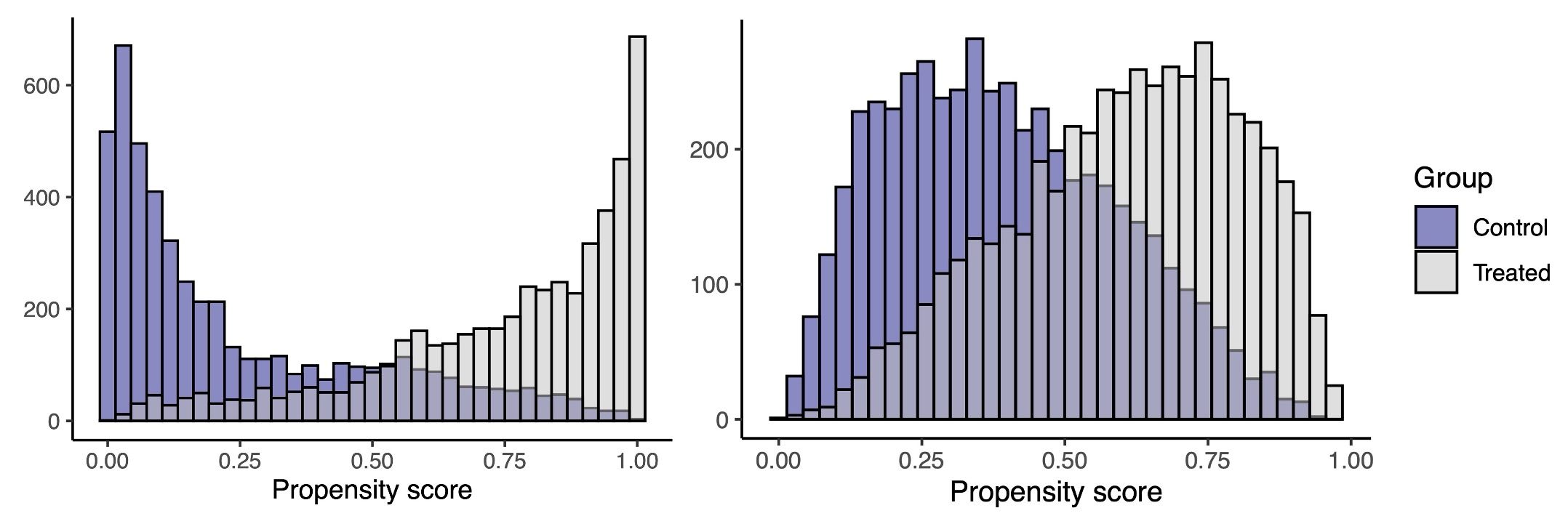}
    \caption{Propensity score distributions of two treatment groups under the poor overlap of DGP I (left) and DGP II (right), using a random sample of $N=10^4$. }\label{fig:ps_main}
\end{figure}

\subsection{Results}\label{subsec:results}
\textcolor{black}{In this section, to be succinct and concise, we only present the results under the cases of (i) poor PS overlap and the higher adjusted $R^2$ of DGP I, and (ii) $N=1000$ of DGP II. Results from additional simulations (good and moderate overlaps, smaller adjusted $R^2$ of DGP I, as well as $N=200$ of DGP II) are reported in Appendix \ref{apx:simu}. The main conclusion by simulation remains the same across different cases. }

Figure \ref{fig:ps_main} shows the PS distributions by treatment groups under \textcolor{black}{the true DGP I (the left panel)} of the ``poor overlap'' PS model and \textcolor{black}{under the true DGP II (the right panel)}. Clearly, in the left panel, the tails of the PS distributions of two treatment groups have very limited overlap and there is a number of observations in the control group with $e(X)\approx 1$, which can incur extreme weights when estimating ATT. In the right panel, we see moderate overlap between PS distributions and there are a smaller number of extreme weights than the left panel. 

\begin{table}[H]
    \singlespacing
    \footnotesize
    \centering
    \begin{tabular}{rrrrrrrrr}
    \toprule
        Method & ARBias\% & RMSE & CP\% & ARBias\% & RMSE & CP\%\\
        \midrule
         & \multicolumn{3}{c}{PS model correctly specified} & \multicolumn{3}{c}{PS model misspecified}\\ 
         \cmidrule(rl){2-4}\cmidrule(rl){5-7}
        ATT & 1.64 & 1.19 & 91.20 & 2.32  & 2.93 & 69.20 \\
        \addlinespace
        \bf{OWATT} & \bf 0.13 & \bf 0.66 & \bf 94.40 & \bf 5.91 & \bf 1.20 & \bf 76.80\\
        \addlinespace
        ATT trimming ($\alpha=0.05$),  PS re-estimated & 0.68 & 0.70 & 94.90 &  6.60 & 1.30 & 70.80  \\
        ATT trimming ($\alpha=0.10$),  PS re-estimated & 1.67 & 0.73 & 94.50 & 7.78 & 1.42 & 62.80 \\
        ATT trimming ($\alpha=0.15$),  PS re-estimated & 2.81 & 0.82 & 91.20 & 9.45 & 1.60  & 49.90 \\
        \addlinespace
     ATT trimming ($\alpha=0.05$), PS not re-estimated & 0.08 & 0.68 & 94.80 & 4.80 & 1.07 &  84.30 \\
        ATT trimming ($\alpha=0.10$), PS not re-estimated & 0.18 & 0.68 & 94.90  & 4.56 & 1.02 & 86.10  \\
         ATT trimming ($\alpha=0.15$), PS not re-estimated & 0.21 & 0.69 & 94.10 & 4.95 & 1.05  &  84.10 \\
       \addlinespace
        Smooth ATT trimming ($\alpha=0.05, \varepsilon=0.001$) & 0.09 & 0.68 & 95.00 & 4.80 & 1.07 & 84.10 \\
        Smooth ATT trimming ($\alpha=0.10, \varepsilon=0.001$) & 0.18 & 0.68 & 94.90 & 4.55 & 1.02 & 86.10 \\
        Smooth ATT trimming ($\alpha=0.15, \varepsilon=0.001$) & 0.21 & 0.69 & 94.20  & 4.95 & 1.05 & 84.10 \\
        \addlinespace
        Smooth ATT trimming ($\alpha=0.05, \varepsilon=0.01$) & 0.10 & 0.68 & 94.90 & 4.84 & 1.07 & 83.30 \\ 
  Smooth ATT trimming ($\alpha=0.10, \varepsilon=0.01$) & 0.18 & 0.68 & 94.70 & 4.57 & 1.02 & 85.80 \\ 
  Smooth ATT trimming ($\alpha=0.15, \varepsilon=0.01$) & 0.20 & 0.69 & 94.40 & 4.97 & 1.05 & 83.90 \\ 
        \addlinespace
        Smooth ATT trimming ($\alpha=0.05, \varepsilon=0.05$) & 0.13 & 0.74 & 94.70 & 3.67 & 1.58 & 80.60 \\ 
  Smooth ATT trimming ($\alpha=0.10, \varepsilon=0.05$) & 0.11 & 0.66 & 94.40 & 4.63 & 1.08 & 82.80 \\ 
  Smooth ATT trimming ($\alpha=0.15, \varepsilon=0.05$) & 0.17 & 0.66 & 94.20 & 5.11 & 1.06 & 83.20 \\
        \addlinespace
        ATT truncation ($\alpha=0.05$) & 0.11 & 0.70 & 94.30 & 5.09 & 1.14 & 80.70 \\
        ATT truncation ($\alpha=0.10$) & 0.12 & 0.68 & 94.20 & 4.95 & 1.09 & 82.30  \\ 
        ATT truncation ($\alpha=0.15$) & 0.13 & 0.67 & 93.90 & 4.96 & 1.07 & 82.80 \\
       \bottomrule
    \end{tabular}
    \begin{tablenotes}\footnotesize
        \item PS: propensity score; ARBias\%: absolute relative bias (\%); {RMSE: root mean square error;} CP\%: coverage probability (\%); ATT: average treatment effect on the treated; OWATT: overlap weighted average treatment effect on the treated. 
    \end{tablenotes}
    \caption{Simulation results under poor PS overlap of DGP I}\label{tab:result_main}
\end{table}

\begin{table}[H]
    \singlespacing
    \footnotesize
    \centering
    \textcolor{black}{
    \begin{tabular}{rrrrrrrrr}
    \toprule
        Method & ARBias\% & RMSE & CP\% & ARBias\% & RMSE & CP\%\\
        \midrule
         & \multicolumn{3}{c}{PS model correctly specified} & \multicolumn{3}{c}{PS model misspecified}\\ 
         \cmidrule(rl){2-4}\cmidrule(rl){5-7}
        ATT & 0.12 & 2.50 & 92.80 & 35.63  & 7.32 & 3.40 \\
        \addlinespace
        \bf{OWATT} & \bf 0.31 & \bf 1.22 & \bf 95.30 & \bf 24.83 & \bf 4.43 & \bf 20.10\\
        \addlinespace
        ATT trimming ($\alpha=0.05$),  PS re-estimated & 0.31 & 1.80 & 94.10 & 34.09 & 6.84 & 3.90  \\
        ATT trimming ($\alpha=0.10$),  PS re-estimated & 1.67 & 1.64 & 95.20 & 28.90 & 5.37 & 13.60 \\
        ATT trimming ($\alpha=0.15$),  PS re-estimated & 6.30 & 1.82 & 91.40 & 21.73 & 3.73  & 49.40 \\
        \addlinespace
        ATT trimming ($\alpha=0.05$), PS not re-estimated & 0.76 & 1.81 & 94.20 & 34.09 & 6.84 & 3.80 \\
        ATT trimming ($\alpha=0.10$), PS not re-estimated & 1.00 & 1.68 & 95.70 & 28.90 & 5.37 & 12.80  \\
         ATT trimming ($\alpha=0.15$), PS not re-estimated & 0.57 & 1.63 & 96.50 & 21.55 & 3.69 & 48.40 \\
       \addlinespace
        Smooth ATT trimming ($\alpha=0.05, \varepsilon=0.001$) & 0.77 & 1.80 & 94.10 & 34.10 & 6.83 & 3.90 \\
        Smooth ATT trimming ($\alpha=0.10, \varepsilon=0.001$) & 0.97 & 1.66 & 95.60 & 28.90 & 5.37 & 12.50 \\
        Smooth ATT trimming ($\alpha=0.15, \varepsilon=0.001$) & 0.60 & 1.62 & 96.30 & 21.55 & 3.69 & 48.20 \\
        \addlinespace
        Smooth ATT trimming ($\alpha=0.05, \varepsilon=0.01$) & 0.59 & 1.73 & 94.10 & 34.00 & 6.81 & 3.60 \\ 
        Smooth ATT trimming ($\alpha=0.10, \varepsilon=0.01$) & 0.91 & 1.59 & 95.00 & 28.91 & 5.36 & 11.50 \\ 
        Smooth ATT trimming ($\alpha=0.15, \varepsilon=0.01$) & 0.67 & 1.57 & 95.80 & 21.66 & 3.70 & 46.40 \\ 
        \addlinespace
        Smooth ATT trimming ($\alpha=0.05, \varepsilon=0.05$) & 0.31 & 1.62 & 94.20 & 32.90 & 6.46 & 3.40 \\ 
        Smooth ATT trimming ($\alpha=0.10, \varepsilon=0.05$) & 0.57 & 1.36 & 95.60 & 28.79 & 5.29 & 10.40 \\ 
        Smooth ATT trimming ($\alpha=0.15, \varepsilon=0.05$) & 0.64 & 1.37 & 95.60 & 22.55 & 3.82 & 36.90 \\
        \addlinespace
        ATT truncation ($\alpha=0.05$) & 0.18 & 2.02 & 93.30 & 35.17 & 7.18 & 3.00 \\
        ATT truncation ($\alpha=0.10$) & 0.42 & 1.65 & 94.30 & 33.41 & 6.63 & 3.60  \\ 
        ATT truncation ($\alpha=0.15$) & 0.48 & 1.43 & 95.10 & 30.87 & 5.87 & 6.20 \\
       \bottomrule
    \end{tabular}
    \begin{tablenotes}\footnotesize
        \item PS: propensity score; ARBias\%: absolute relative bias (\%); {RMSE: root mean square error;} CP\%: coverage probability (\%); ATT: average treatment effect on the treated; OWATT: overlap weighted average treatment effect on the treated. 
    \end{tablenotes}}
    \caption{Simulation results under DGP II with $N=1000$}\label{tab:result_main_II}
\end{table}

{\color{black}In Tables \ref{tab:result_main} and \ref{tab:result_main_II}, we summarized the results of the cases we chose to report in this section, under both correctly specified and misspecified PS models. In both tables, when the PS model is correctly specified, we see that the OWATT estimator has the smallest RMSE (0.66 by OWATT vs. [0.66, 1.19] by the other methods in Table \ref{tab:result_main}, and 1.22 by OWATT vs. [1.36, 2.50] by the other methods in Table \ref{tab:result_main_II}). The OWATT estimator also has valid CP\% close to the 95\% nominal level when the PS model is correctly specified. These indicate that among methods we assessed, OWATT has the highest internal validity and is efficient under lack of positivity when the PS is estimated well. 

When the PS model is misspecified, all methods are biased in both tables, but this is expected because all methods are dependent to PS. However, this gives us a chance to compare the sensitivity of different methods under the PS model misspecification, which is the main purpose of this simulation design. As can be seen, in this case, OWATT still outperforms than the conventional ATT and most of other ad hoc methods, showing smaller RMSE values. Evidently, in both tables, the ATT estimator performs the worst in both PS model specifications, which is expected because the efficiency of the simple weighting estimator is affected by the extreme weights from control participants whose PS $e(X)$ that are near or equal to 1 (see Figure \ref{fig:ps_main}). These findings are also consistent with our conclusion in Theorem \ref{th:watt}.
}

%In Table \ref{tab:result_main}, the OWATT estimator has smaller amount of variation on point estimation and the mean of the relative biases is one of the closest to 0. When the PS model is correctly specified, it has the smallest RMSE (0.66), compared to RMSEs of other methods ranging in $[0.66, 1.19]$. OWATT estimator is also one of the most robust methods to PS model misspecification, especially when compared to ATT it terms of their RMSEs (1.20 vs. 2.93).  
In addition, ATT trimming with PS not re-estimated (for $\alpha=0.05, 0.10$ and $0.15$) and smooth ATT trimming with the $\varepsilon=0.001$ or $0.01$ (for $\alpha=0.05, 0.10$ and $0.15$) perform similarly to OWATT, under correctly specified PS model.  When $\varepsilon=0.05$,  the RMSE of smooth ATT trimming can be slightly larger (e.g., when $(\alpha,\varepsilon)=(0.05,0.05)$). Thus, with larger $\varepsilon$'s, the performance of smooth ATT trimming is sensitive to the choice of $\alpha$. 

Overall, OWATT has good performances in both cases of the PS model specification. ATT trimming, smooth or not,  does not indicate a clear and consistent pattern when we increase $\alpha$, $\varepsilon$, or both. Our simulation study also shows that re-estimating PSs has a larger bias when PS model is correctly specified or when it is misspecified. 

As indicated, we have included all the other simulation results in Appendix \ref{apx:simu} for further details, but we briefly include some summary here to showcase the advantage of our method. Figure \ref{fig:ps-distns} shows the PS plots of all 3 PS models {\color{black} under DGP I.} Figure \ref{fig:Rbias-B.1} shows the boxplots of relative biases under all the 3 overlaps \textcolor{black}{and the case of smaller adjusted $R^2$ under DGP I. Figure \ref{fig:Rbias-Kang} shows the boxplots of relative biases under DGP II with both sample sizes ($N=200$ and $1000$). } These boxplots help us better understand the variations of relative biases among multiple Monte Carlo replicates we considered, especially when these relative biases are skewed. For example, based on the results from both Tables \ref{tab:result_main} and \ref{tab:res_poor} under poor overlap \textcolor{black}{under DGP I}, the mean bias for the conventional ATT estimator under model misspecification is smaller, but from Figure \ref{fig:Rbias-B.1}, we can see there are a number of extremely large biases in some of the  data replicates. Thus, the distribution of these biases is expected to be skewed. In addition, from Figure \ref{fig:Rbias-B.1}, we can see in both model specification cases, the OWATT performs well as one of the most stable methods (regarding the variation of relative bias), while some ad-hoc methods do not show a consistent pattern by changing the threshold (and possibly with a smooth parameter), e.g., truncation and smooth ATT trimming with $\varepsilon=0.01$. 

\textcolor{black}{Tables \ref{tab:res_good}--\ref{tab:res_poor-smallR2} present the complete results under good, moderate, poor overlaps and smaller $R^2$ of DGP I under $N=1000$, respectively. Tables \ref{tab:res-kang-n200} and \ref{tab:res-kang-n1000} present the complete results of DGP II under $N=200$ and $N=1000$, respectively. From all these results, } we found that they are similar to our findings above in Tables \ref{tab:result_main} and \ref{tab:result_main_II}. However, it can be observed that in the good and moderate overlap cases \textcolor{black}{under DGP I}, the differences in performances of these methods are overall smaller due to the increasing in overlap. \textcolor{black}{For DGP II under $N=200$, we observed larger RMSEs by all methods due to the samller sample size compared to $N=1000$, but the conclusion remains unchanged that the OWATT estimator is the most efficient method. }

\section{Data Application}\label{sec:data}

To illustrate our methods, we conducted a data analysis on racial disparities in health care expenditure using data from the Medical Expenditure Panel Survey (MEPS) (\url{ https://www.meps.ahrq.gov/mepsweb/}). Our analysis included 11276 individuals, with 9830 (87.18\%) non-Hispanic White and 1446 (12.82\%) Asian people, where White is considered as the treated group ($Z=1$) and Asian as the control group ($Z=0$). We considered in a total of 31 covariates, including 4 continuous and 27 binary variables and  (individual ) health care expenditure as the outcome of interest. 

Note that we use the terms ``treated'' and ``controls'' in this Section \ref{sec:data} to match the language and the notations of the previous sections. However, since racial/ethnic membership is not manipulable, the PS weighting methods presented in this paper are being leveraged to simply balance the distributions of the covariates between the racial groups and better assess racial disparities in health care utilization. { Such a comparison is also called controlled descriptive comparison.  \cite{li2023using} In this context, the ATT is a meaningful measure for racial disparities. It evaluates the  average difference in health care expenditure between White and Asian participants had the covariate distribution of Asian participants were made (or forced to be) similar to that of White participants (our reference group).\cite{li2023using} }

\begin{figure}[H]
    \centering   \includegraphics[trim=10 05 10 05, clip, width=0.8\textwidth]{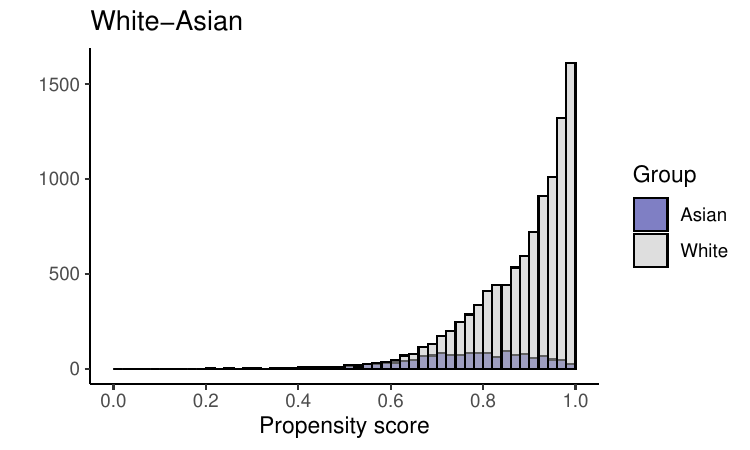}
    \caption{Propensity score distributions of White vs. Asian in MEPS data}
    \label{fig:ps-meps}
\end{figure}

Figure \ref{fig:ps-meps} shows the distributions of the estimated PS of the two race groups, which {was} estimated using a multivariable logistic regression model on all 31 covariates. Overall, the PS distributions are left-skewed and have good overlap. Most Asian participants have PSs greater than 0.6 and a substantial number of them have PSs near 1. This suggests that certain Asian participants will have large conventional ATT weights in comparison. In addition, there can be a substantial loss in sample size if we decide to trim control (i.e., Asian) participants with large PS weights. Indeed, when we trim at the threshold of $\alpha=0.05$,  1356 (93.78\%) participants remain; at $\alpha=0.10$, the number of Asian participants decreases to 1202 (83.13\%); and if we trim at $\alpha=0.15$, only 995 (68.81\%) participants remain in the group.

\begin{figure}[H]
    \centering
    \includegraphics[trim=05 05 05 05, clip, width=0.9\textwidth]{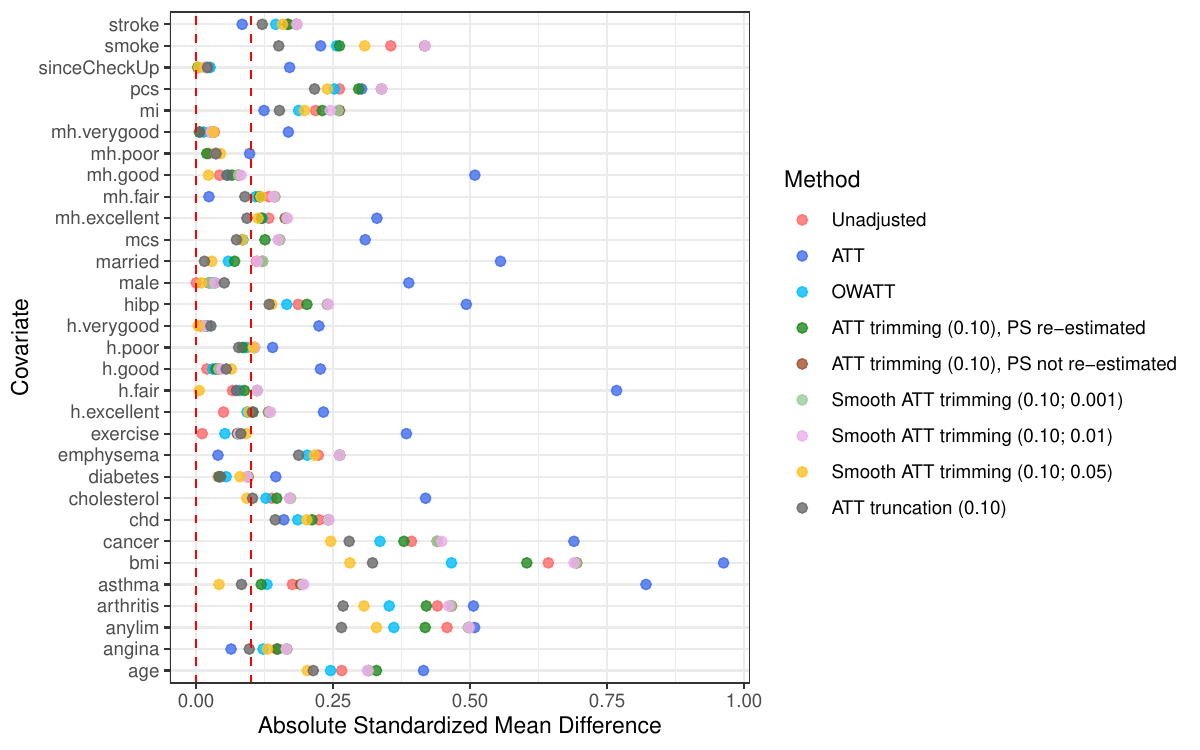}
    \begin{tablenotes}\footnotesize
    \item *The two red vertical dashed lines indicate the absolute difference within $\pm 0.10$.  
    \end{tablenotes}
    \caption{Covariates balance of White vs. Asian in MEPS data}
    \label{fig:love-meps-part}
\end{figure}

Figure \ref{fig:love-meps-part} presents the covariates balance by 9 different estimand weights, measured by {absolute} standardized mean difference (ASD) of each covariate between the two groups. The ASD of a single covariate $x_j$, $j=1,\dots,31$, is defined by \cite{austin2015moving}
\begin{align*}
    {d_j = \left\vert\dfrac{\bar x_{j0}-\bar x_{j1}}{\sqrt{(s^2_{j1}+s_{j0}^2)/2}}\right\vert},~~\text{with}~\bar x_{jz} = \left\{\displaystyle\sum_{i=1}^N W_{z,i}\right\}^{-1}\displaystyle\sum_{i=1}^N W_{z,i}x_{j,i}
\end{align*}
where $\bar x_{jz}$ is the weighted mean of the covariate $x_j$ over the subgroup $z=0$ or $1$, and $$
s_{jz}^2 = \dfrac{\displaystyle\sum_{i=1}^N W_{z,i}}{\left(\displaystyle\sum_{i=1}^N W_{z,i}\right)^2 - \displaystyle\sum_{i=1}^N W_{z,i}^2}\displaystyle\sum_{i=1}^N W_{z,i}(x_{j,i}-\bar x_{jz})^2,~~z=0,1
$$
where $W_{1,i} = Z_i$ and $W_{0,i} = (1-Z_i)\widehat\omega_{0h}(X_i)$. {It can be seen that for the {conventional}  ATT weights, several ASDs exceed the $0.1$ threshold more frequently and some of  them are the largest across all the methods we considered}. In general, OWATT, smooth ATT trimming ($\alpha=0.10, \varepsilon=0.05$), and ATT truncation ($\alpha=0.10$) balance the covariates the best. {Looking at both Figures \ref{fig:ps-meps} and \ref{fig:love-meps-part}, we can see that because extreme conventional ATT  weights exist for a few observations in control group (Asian), the covariate balance by conventional ATT  weights is unstable and even worse than those from the unadjusted (i.e., crude, non-weighted) method. }

Table \ref{tab:meps-effects} shows 20 different WATTs, their points estimates and their estimated standard error (SE),  using 500 bootstrap replicates. The OWATT has a point estimate of \$2511.91 and ATT has \$2399.32. Moreover, the point estimate of each ad-hoc method (trimming, smooth trimming, and truncation) increases as the threshold $\alpha$ increases. The estimate from the trimming methods goes from  less than \$2500 to greater \$3200, while those from truncation remain below  \$2500. For example,  with ATT trimming where the PS is not re-estimated, the point estimates are, respectively, \$2363.09, \$2666.13, and \$3054.09 for $\alpha=0.05, 0.10$ and $0.15$, respectively.  Similarly, for a fixed $\alpha$, the estimate of the smooth ATT trimming increases as $\varepsilon$ decreases.

In terms of estimated bootstrap variances, the OWATT estimator has the smallest SE, except when compared against smooth ATT trimming for $(\alpha,\varepsilon)=(0.10,0.05)$ and $(0.15,0.05)$. As expected, the ATT estimator has the largest SE due to the presence of extreme weights.  For both trimming with and without re-estimating the PS, the higher the threshold $\alpha$, the lower the standard error. With smooth trimming,  for a fixed $\alpha$, an increase in $\varepsilon$, results also in a decrease in SE, except when $(\alpha,\varepsilon)$ goes from $(0.05,0.01)$ to $(0.05,0.05)$. For the same value of $\alpha$, the SE for truncation is better only when compared to ATT trimming without PS re-estimation. Otherwise, truncation is less efficient than ATT trimming with PS not re-estimated and smooth ATT trimming for the same threshold $\alpha$ Overall, the OWATT estimator provides the best bias-variance trade-off and is better than truncation, both in teasing out the point estimate and by improving efficiency. Both trimming (smooth or not) and truncation give the impression that the more we trim or truncate, the better off we are. However, as demonstrated through our simulation studies, this can be misleading and may lead to a fishing expedition on where to stop in the choice of $\alpha.$

\begin{table}
    \singlespacing
    \centering
    \small
    \begin{tabular}{rrrrr}
      \toprule
    Method & Point estimate & Standard error & p-value \\ 
      \midrule
    ATT & 2399.32 & 787.37 & 0.002 \\ 
    \addlinespace
     \bf  OWATT & \bf 2511.91 & \bf 255.20 & $<0.001$ \\ 
      \addlinespace
      ATT trimming ($\alpha=0.05$), PS re-estimated & 2363.09 & 403.42 & $<0.001$ \\ 
      ATT trimming ($\alpha=0.10$), PS re-estimated & 2666.13 & 356.62 & $<0.001$ \\ 
      ATT trimming ($\alpha=0.15$), PS re-estimated & 3054.09 & 352.98 & $<0.001$ \\  
      \addlinespace
      ATT trimming ($\alpha=0.05$), PS not re-estimated & 2487.25 & 352.16 & $<0.001$ \\ 
      ATT trimming ($\alpha=0.10$), PS not re-estimated & 2928.39 & 286.52 & $<0.001$ \\ 
      ATT trimming ($\alpha=0.15$), PS not re-estimated & 3286.90 & 270.04 & $<0.001$ \\ 
      \addlinespace
      Smooth ATT trimming ($\alpha=0.05, \varepsilon=0.001$) & 2488.98 & 348.88 & $<0.001$ \\ 
      Smooth ATT trimming ($\alpha=0.10, \varepsilon=0.001$) & 2926.52 & 285.92 & $<0.001$ \\ 
      Smooth ATT trimming ($\alpha=0.15, \varepsilon=0.001$) & 3291.05 & 268.68 & $<0.001$ \\ 
      \addlinespace
      Smooth ATT trimming ($\alpha=0.05, \varepsilon=0.01$) & 2419.59 & 327.68 & $<0.001$ \\ 
      Smooth ATT trimming ($\alpha=0.10, \varepsilon=0.01$) & 2881.88 & 277.57 & $<0.001$ \\ 
      Smooth ATT trimming ($\alpha=0.15, \varepsilon=0.01$) & 3229.41 & 259.47 & $<0.001$ \\ 
      \addlinespace
      Smooth ATT trimming ($\alpha=0.05, \varepsilon=0.05$) & 2337.55 & 373.65 & $<0.001$ \\ 
      Smooth ATT trimming ($\alpha=0.10, \varepsilon=0.05$) & 2638.19 & 250.78 & $<0.001$ \\ 
      Smooth ATT trimming ($\alpha=0.15, \varepsilon=0.05$) & 3014.23 & 232.06 & $<0.001$ \\ 
      \addlinespace
      ATT truncation ($\alpha=0.05$) & 1945.35 & 385.00 & $<0.001$ \\ 
      ATT truncation ($\alpha=0.10$) & 2211.56 & 307.63 & $<0.001$ \\ 
      ATT truncation ($\alpha=0.15$) & 2419.23 & 271.39 & $<0.001$ \\ 
      \bottomrule
    \end{tabular}   
    \caption{Estimated WATTs for health care expenditure among White vs. Asian from MEPS data}
    \label{tab:meps-effects}
\end{table}

\section{Discussion}\label{sec:remarks}

The positivity assumption is crucial to any propensity score (PS) method, but it is formulated differently depending on the estimand of interest and the method considered. For the average treatment effect on the treated (ATT), the postivity assumption requires that the PS of control participants be strictly less than 1. Violation (or near violations) of the positivity assumption leads to biased and even inefficient estimate of the ATT.  Faced with this issue, the common practice is to target an ATT-like estimand by choosing an interval $[0, 1-\alpha]$  and either (1) trim i.e., exclude (smoothly or not) control participants whose PS falls outside the interval or (2) cap their weights to a fixed value (e.g., $\alpha(1-\alpha)^{-1}$), for a user-selected threshold $\alpha$. \cite{lee2011weight, cole2008constructing}

In this paper, we circumvented the subjectivity of selecting such a user-specified threshold $\alpha$ (as well as the smooth parameter $\varepsilon$)  and considered instead a data-driven mechanism by multiplying the tilting function $h(x)=e(x)\{1-e(x)\}$ of the overlap weight to the conventional ATT  weights $w_0(x) = e(x)/\{1-e(x)\}$ of control participants. We have demonstrated that using such a tilting function has several advantages. First, it helps fulfill the positivity assumption, targets the population of control participants for whom there is a clinical equipoise,\cite{li2018balancing, matsouaka2020framework} and leads to efficient estimation. Second, the OWATT it helps characterize the treatment effect better  when there is treatment effect heterogeneity. Indeed, using different thresholds (and possibly some smoothing parameters) for ATT trimming, smooth ATT trimming or ATT truncation can result in inconsistent pattern of treatment effect estimates when the effect is heterogeneous, as illustrated in our simulation and data analysis. On the other hand, OWATT uses available information from all participants, weighting each participant adequately based on their PS. {Therefore, OWATT can potentially be a better alternative to ATT, under violations of positivity assumption---especially when such a violation is structural, i.e., due to inherent characteristics of the target population. \cite{petersen2012diagnosing, westreich2010invited, matsouaka2020framework} 

In addition, the proposed method is flexible and can be leveraged for any other appropriate $h(x)$ functions for WATT estimand, which also helps define the target population $\mc H$. We plan to investigate in our future empirical studies the use of other $h(x)$, such as the Shannon's entropy function or the matching function, that currently available in literature. \cite{zhou2020propensity, li2013weighting, matsouaka2020framework, matsouaka2024overlap}  Moreover, our framework can be easily extended to the average treatment effect on the control (ATC), which is also considered in assessing the {positive effect} of interventions that aim to curb or end nefarious exposures. {This can be, for example,} an intervention for smoking cessation {in a population of young adults} or  {youth} drug rehabilitation programs. Such an investigation has a considerable practical implication, especially when it comes to policy decision-making. \cite{matsouaka2023variance} When there is a lack of positivity in estimating ATC, we suggest the  use of  $\omega_{0h}(x) = (1-e(x))^2$ to appropriately weight the treated participants and estimate the weighted ATC (WATC) as an alternative. 

As shown in our simulation study, the methods presented in this paper are all biased when the PS model is misspecified, particularly the {conventional} ATT (without trimming and without truncation) weighting estimator which is the least robust against model misspecifications. Therefore, in practice, we suggest that a PS model as saturated as possible be postulated to gain higher accuracy of prediction. In that regard, one can use techniques such as penalized regression, series estimation, \cite{newey1994series} or machine learning methods.\cite{lee2010improving, li2021propensity} In addition, we would like to mention that using machine learning methods may improve the PS estimation, and thus avoid the violation of positivity due to the misspecification of the parametric model. However, this does not help in the case where there is a structural violation of positivity assumption when the true PS distribution at the population level lacks  positivity (as considered in our simulation in Section \ref{sec:sim}, too). 
\cite{petersen2012diagnosing} {A similar comment is about the use of another popular weighting method in causal inference, the calibration weighting (CW).\cite{imai2014covariate, su2022sensitivity, kott2016calibration} CW used weights calculated based on some finite-sample moment equations, which also imposes covariate balance between the treatment groups. It can be viewed as another way to estimate the PS weights using some covariate balancing constraints, instead of the maximum likelihood estimation (MLE). However, when the violation of the positivity assumption is structural, the true covariate distributions between the treatment groups can be systematically different. Therefore, in this case,} we suggest to ``move the goalpost'' to some subpopulation where the positivity is satisfied.\cite{crump2006moving} In the context of ATT, the goalpost is a weighted or sub-population of the control participants. Our proposal of OWATT is a special case of such goalpost on controls, which can deal with both random and structural violations of positivity. 

Finally, {there are a number of important aspects of the WATT for which  we leave a thorough
investigation to future work. First,} we have not investigated the semiparametric efficiency of the WATT estimators, \cite{abadie2005semiparametric, crump2009dealing, hirano2003efficient} which can include additional modeling on outcomes and the study of corresponding augmented estimators. Furthermore, we can also leverage the M-theory,  \cite{stefanski2002calculus, tsiatis2007semiparametric} and build sandwich-variance estimators \cite{lunceford2004stratification} for the aforementioned weighting and augmented estimators and, ultimately, assess their robustness as well as their efficiency. {Lastly, we have demonstrated in this paper that OWATT is a better alternative than {the conventional} ATT under lack of positivity. However, it is also worthy to investigate in theory and {with great} details how similar is OWATT to ATT when the overlap is good or moderate, which could potentially make the OWATT a more widely considered tool in more cases.  }

\begin{acks}
The authors are grateful to Dezhao Fu (Master of Biostatistics, Duke University) and Chenyin Gao (Department of Statistics, North Carolina State University), for helpful discussions in early stage of this research endeavor. We also thank Shi Cen (Bioinformatics Research Center, North Carolina State University) for his computational support. \textcolor{black}{We especially thank the two anonymous reviewers and the editor for their patient and constructive feedback, which improved tremendously the quality of this paper. }
\end{acks}
	
\newpage
\bibliographystyle{SageV}
\bibliography{OWATT}

\newpage
\appendix
\newcounter{Appendix}[section]
\numberwithin{equation}{subsection}
\renewcommand\theequation{\Alph{section}.\arabic{subsection}.\arabic{equation}}
\numberwithin{table}{subsection}
\numberwithin{figure}{subsection}
\section{Appendix: Technical Proofs}\label{apx:tech}

\subsection{Equivalent formulas for ATT}\label{subapx:equi-form-att}

Denote $e(X) = P(Z=1\mid X)$, we have
\allowdisplaybreaks \begin{align*}
    \tau_{att} & = \Ex\{Y(1)-Y(0)\mid Z=1\} = \Ex\{Y\mid Z=1\} - \Ex\{Y(0)\mid Z=1\} \\
    & = \Ex\{Y\mid Z=1\} - \Ex\{\Ex\{Y(0)\mid X, Z=1\}\mid Z=1\}\\
    & = \Ex\{Y\mid Z=1\} - \Ex\{\Ex\{Y\mid X, Z=0\}\mid Z=1\} ~(\text{unconfoundness})\\
    & = {\Ex(Z)}^{-1}\left[\Ex(ZY) - \Ex\{Z\Ex\{Y\mid X, Z=0\}\}\right]
    \end{align*}
    Furthermore,
 \allowdisplaybreaks    \begin{align*}
    \Ex\{Z\Ex\{Y\mid X, Z=0\}\}& = {\Ex\{\Ex\{Z\Ex\{Y\mid X, Z=0\}\mid X\}\}}\\
    & = {\Ex\{P(Z=1\mid X)\Ex\{Y\mid X,Z=0\}\}}\\
    &=\Ex\left\{P(Z=1\mid X){P(Z=0\mid X)}^{-1}\Ex\{(1-Z)Y\mid X\}\right\}\\
    & =\Ex\left\{{e(X)}{(1-e(X))}^{-1}\Ex\{(1-Z)Y\mid X\}\right\}    
    \end{align*}
    Hence,
 \allowdisplaybreaks    \begin{align*}
  \tau_{att}   & = {\Ex(Z)}^{-1}\left[{\Ex(ZY)}  - \Ex\left\{{e(X)}{(1-e(X))}^{-1}(1-Z)Y\right\}\right]\\
    & = \frac{\Ex(ZY)}{\Ex(Z)}  - \frac{\Ex\left\{w_0(X)(1-Z)Y\right\}}{\Ex\{w_0(X)(1-Z)\}},~\text{with } ~ w_0(X)={e(X)}(1-e(X))^{-1}
\end{align*}
where $\Ex(Z) = P(Z=1) = \Ex\{e(X)\} = \Ex\{w_0(X)(1-Z)\}$  under SUTVA and unconfoundness. This shows the formula \eqref{eq:att} in Section \ref{subsec:framework} in the paper.

\subsection{Proof of Theorem \ref{th:watt}}\label{subapx:th1-pf}
Denote $\widehat\beta_N$ the maximum likelihood estimator (MLE) of $\beta$ in the propensity score (PS) model $e(X)=e(X'\beta^*)$. Define $f(t) = de(t)/dt$ and for given data $\{(X_i, Z_i)\}_{i=1}^N$,
\allowdisplaybreaks \begin{align*}
    S(\beta^*) & = \frac{1}{N}\sum_{i=1}^N\frac{\partial\log L(\beta^*; X_i,Z_i)}{\partial\beta'} = 
    \frac{1}{N}\sum_{i=1}^N X_i\frac{Z_i-e(X_i'\beta^*)}{e(X_i'\beta^*)\{1-e(X_i'\beta^*)\}}f(X_i'\beta^*)\quad(\text{Score function}) \\
    \mathcal{I}(\beta^*) & = \Ex\left\{\frac{f(X'\beta^*)}{e(X'\beta^*)\{1-e(X'
    \beta^*)\}}XX'\right\} \quad(\text{Fisher's information matrix})
\end{align*}
We assume that $h(X)$, as a function of $e(X)$, is continuous almost everywhere with respect to $e(X)$. %,  i.e., $\{X: h(X)\text{ is continuous with respect to } e(X)\}$ is a probability 1 set. 
By Taylor expansion, we have
\allowdisplaybreaks  \begin{align*}
    \widehat\tau_{watt}^h(\mc H) & = \widehat\tau_{watt}^h(\mc H)(\widehat\beta_N) \\
    & = \widehat\tau_{watt}^h(\mc H)(\beta^*) + \Ex\left\{\frac{\partial\widehat\tau_{watt}^h(\mc H)(\beta^*)}{\partial\beta'}\right\}(\widehat\beta_N-\beta^*) + o_p(N^{-1/2}) \\
    & = \frac{1}{N}\sum_{i=1}^N\left\{\frac{Z_i}{\Ex(Z)} - \frac{\omega_{0h}(X'\beta^*)(1-Z)}{\Ex\{\omega_{0h}(X'\beta^*)(1-Z)\}}\right\}Y_i + \Ex\left\{\frac{\partial\widehat\tau_{watt}^h(\mc H)(\beta^*)}{\partial\beta'}\right\}(\widehat\beta_N-\beta^*) + o_p(N^{-1/2})\\
    & = \frac{1}{N}\sum_{i=1}^N\left\{\frac{Z_iY_i}{\Ex\{e(X'\beta^*)\}} - \frac{\omega_{0h}(X_i'\beta^*)(1-Z_i)Y_i}{\Ex\{e(X'\beta^*)h(X'\beta^*)\}}\right\}+ B'\frac{1}{N}\sum_{i=1}^N X_i\frac{Z_i-e(X_i'\beta^*)}{e(X_i'\beta^*)\{1-e(X_i'\beta^*)\}}f(X_i'\beta^*) \\
    & \quad+ o_p(N^{-1/2}),
\end{align*}
where $\omega_{0h}(x)=\dfrac{e(x)h(x)}{1-e(x)}$ and 
$\displaystyle
B' = \Ex\left\{\frac{\partial\widehat\tau_{watt}^h(\mc H)(\beta^*)}{\partial\beta'}\right\}\mathcal{I}(\beta^*)^{-1}. $ \\
The third ``$=$'' follows from $\widehat\beta_N-\beta^* = \mathcal{I}(\beta^*)^{-1}S(\beta^*) + o_p(N^{-1/2})$, under regularity conditions. \citep{van2000asymptotic} Thus, $\widehat\tau_{watt}^h(\mc H)$ is asymptotically linear. 

Moreover, let us denote
\allowdisplaybreaks \begin{align*}
    %m^{(1)}\{e(X'\beta^*)\} & = \Ex\{Y(1)\mid e(X'\beta^*)\},\\
    %m^{(0)}\{e(X'\beta^*)\} & = \Ex\{Y(0)\mid e(X'\beta^*)h(X'\beta^*)\},\\
    \mu(Z,X) = \Ex\{Y\mid X, Z\} ~~\text{and}~~
    \mu\{Z,e(X'\beta^*)\}  = \Ex\{Y\mid e(X'\beta^*), Z\}. 
\end{align*}
We can verify that 
    $\displaystyle N^{1/2}(\widehat\tau_{watt}^h(\mc H)-\tau_{watt}^h(\mc H)) = T_1+T_2+T_3+O_p(N^{-1/2})$ where
\allowdisplaybreaks \begin{align*}
    % T_0 & = N^{-1/2}\sum_{i=1}^N\left\{\frac{m^{(1)}\{e(X'\beta^*)\}}{\Ex\{e(X'\beta^*)\}}-\frac{m^{(0)}\{e(X'\beta^*)\}}{\Ex\{e(X'\beta^*)h(X'\beta^*)\}}-\tau_{watt}\right\},\\
    %%%%%%%%%%%%%%%%%%%%%%%%%%%%%%%%%%%%%%%%%%%%%%%%%%%%%%%%%%%%%%%%%%%%%%%%%%
    %%%%%%%%%%%%%%%%%%%%%%%%%%%%%%%%%%%%%%%%%%%%%%%%%%%%%%%%%%%%%%%%%%%%%%%%%%
    T_1 & = N^{-1/2}\sum_{i=1}^N \bigg\{\frac{Z_i\mu\{Z_i,e(X_i'\beta^*)\}}{\Ex\{e(X'\beta^*)\}} - \frac{\omega_{0h}(X_i'\beta^*)(1-Z_i)\mu\{Z_i,e(X_i'\beta^*)\}}{\Ex\{e(X'\beta^*)h(X'\beta^*)\}} - \tau_{watt}^h(\mc H)\bigg\}\\
    & \quad + N^{-1/2}\sum_{i=1}^N B'\Ex\{X_i\mid e(X_i'\beta^*)\}\frac{Z_i-e(X_i'\beta^*)}{e(X_i'\beta^*)\{1-e(X_i'\beta^*)\}}f(X_i'\beta^*),\\
    %%%%%%%%%%%%%%%%%%%%%%%%%%%%%%%%%%%%%%%%%%%%%%%%%%%%%%%%%%%%%%%%%%%%%%%%%%
    %%%%%%%%%%%%%%%%%%%%%%%%%%%%%%%%%%%%%%%%%%%%%%%%%%%%%%%%%%%%%%%%%%%%%%%%%%
    T_2 & = N^{-1/2}\sum_{i=1}^N  \{\mu(Z_i,X_i)-\mu\{Z_i,e(X_i'\beta^*)\}\}
    \bigg\{\frac{Z_i}{\Ex\{e(X'\beta^*)\}} - \frac{\omega_{0h}(X_i'\beta^*)(1-Z_i)}{\Ex\{e(X'\beta^*)h(X'\beta^*)\}}\bigg\}\\
    & \quad + N^{-1/2}\sum_{i=1}^N B'\{X_i-\Ex\{X_i\mid e(X_i'\beta^*)\}\}\frac{Z_i-e(X_i'\beta^*)}{e(X_i'\beta^*)\{1-e(X_i'\beta^*)\}}f(X_i'\beta^*),\\
    %%%%%%%%%%%%%%%%%%%%%%%%%%%%%%%%%%%%%%%%%%%%%%%%%%%%%%%%%%%%%%%%%%%%%%%%%%
    %%%%%%%%%%%%%%%%%%%%%%%%%%%%%%%%%%%%%%%%%%%%%%%%%%%%%%%%%%%%%%%%%%%%%%%%%%
    T_3 & = N^{-1/2}\sum_{i=1}^N \left\{\frac{Z_i\{Y_i-\mu(Z_i,X_i)\}}{\Ex\{e(X'\beta^*)\}} + \frac{\omega_{0h}(X'\beta^*)(1-Z_i)\{Y_i-\mu(Z_i,X_i)\}}{\Ex\{e(X'\beta^*)h(X'\beta^*)\}}\right\}.
\end{align*}
Furthermore, define 
$\displaystyle\mc F_0 = \{X_1'\beta^*,\dots,X_N'\beta^*\}, ~\mc F_1 = \{Z_1,\dots,Z_N,X_1'\beta^*,\dots,X_N'\beta^*\}$ and $\displaystyle
    \mc F_2 = \{Z_1,\dots,Z_N,X_1'\beta^*,\dots,X_N'\beta^*, X_1,\dots, X_N\}.$
By conditioning arguments, for $k=1,2,3$, we have $\Ex(T_k) = \Ex\{\Ex(T_k\mid \mc F_{k-1})\}=\Ex\{0\} =0$. 

Moreover, for $k=2,3$,
\allowdisplaybreaks \begin{align*}
    \cov(T_1,T_k) & = \cov\{\Ex(T_1\mid \mc F_1), \Ex(T_k\mid \mc F_1)\} + \Ex\{\cov(T_1,T_k\mid \mc F_1)\}\\
    & = \cov\{\Ex(T_1\mid \mc F_1), 0\} + \Ex\{0\} = 0,
\end{align*}
and
\allowdisplaybreaks \begin{align*}
    \cov(T_2,T_3) & = \cov\{\Ex(T_2\mid \mc F_2), \Ex(T_3\mid \mc F_2)\} + \Ex\{\cov(T_2,T_3\mid \mc F_2)\}\\
    & = \cov\{\Ex(T_2\mid \mc F_2), 0\} + \Ex\{0\} = 0.
\end{align*}
Therefore, we have, $\var(T_1+T_2+T_3) = \displaystyle\sum_{k=1}^3\var(T_k) = \displaystyle\sum_{k=1}^3\Ex(T_k^2)$. Specifically,
\allowdisplaybreaks \begin{align*}
    \var(T_1) & = \Ex\{\var\{T_1\mid \mc F_0\}\}\\
    & = \Ex\left\{\frac{\mu\{1,e(X'\beta^*)\}^2e(X'\beta^*)}{\Ex\{e(X'\beta^*)\}^2} + \frac{\mu\{0,e(X'\beta^*)\}^2\omega_{0h}(X'\beta^*)^2\{1-e(X'\beta^*)\}}{\Ex\{e(X'\beta^*)h(X'\beta^*)\}^2}\right\}\\
    & \quad + 2B'\Ex\left\{\Ex\{X\mid e(X'\beta^*)\}\left[\frac{\mu\{1,e(X'\beta^*)\}}{\Ex\{e(X'\beta^*)\}} + \frac{\omega_{0h}(X'\beta^*)\mu\{0,e(X'\beta^*)\}}{\Ex\{e(X'\beta^*)h(X'\beta^*)\}}\right]f(X'\beta^*)\right\}\\
    & \quad + B'\Ex\left\{f(X'\beta^*)^2\frac{\Ex\{X\mid e(X'\beta^*)\}\Ex\{X'\mid e(X'\beta^*)\}}{e(X'\beta^*)\{1-e(X'\beta^*)\}}\right\}B,\\
    \\
    %%%%%%%%%%%%%%%%%%%%%%%%%%%%%%%%%%%%%%%%%%%%%%%%%%%%%%%%%%%%%%%%%%
    %%%%%%%%%%%%%%%%%%%%%%%%%%%%%%%%%%%%%%%%%%%%%%%%%%%%%%%%%%%%%%%%%%
    \var(T_2) & = \Ex\{\var\{T_2\mid \mc F_1\}\}\\
    & = \Ex\left\{\frac{\sigma^2\{1,e(X'\beta^*)\}e(X'\beta^*)}{\Ex\{e(X'\beta^*)\}^2} + 
    \frac{\sigma^2\{0,e(X'\beta^*)\}\omega_{0h}(X'\beta^*)^2\{1-e(X'\beta^*)\}}{\Ex\{e(X'\beta^*)h(X'\beta^*)\}^2}\right\}\\
    & \quad + 2B'\Ex\left\{\left[\frac{\cov\{X,\mu(1,X)\mid e(X'\beta^*)\}}{\Ex\{e(X'\beta^*)\}} + \frac{\cov\{X,\mu(0,X)\mid e(X'\beta^*)\}\omega_{0h}(X'\beta^*)}{\Ex\{e(X'\beta^*)h(X'\beta^*)\}}\right]f(X'\beta^*)\right\} \\
    & \quad + B'\Ex\left\{f(X'\beta^*)^2\frac{\var\{X\mid e(X'\beta^*)\}}{e(X'\beta^*)\{1-e(X'\beta^*)\}}\right\} B,\\
    \\
    %%%%%%%%%%%%%%%%%%%%%%%%%%%%%%%%%%%%%%%%%%%%%%%%%%%%%%%%%%%%%%%%%%
    %%%%%%%%%%%%%%%%%%%%%%%%%%%%%%%%%%%%%%%%%%%%%%%%%%%%%%%%%%%%%%%%%%
    \var(T_3) & = \Ex\{\var\{T_3\mid \mc F_2\}\}\\
    & = \Ex\left\{\sigma^2(1,X)\frac{e(X'\beta^*)}{\Ex\{e(X'\beta^*)\}^2} + \sigma^2(0,X)\frac{\omega_{0h}(X'\beta^*)^2\{1-e(X'\beta^*)\}}{\Ex\{e(X'\beta^*)h(X'\beta^*)\}^2}\right\}.
\end{align*}
In addition, we express $\dfrac{\partial\widehat\tau_{watt}^h(\mc H)(\beta^*)}{\partial\beta'}$ as follows.
\begin{align*}
    \dfrac{\partial\widehat\tau_{watt}^h(\mc H)(\beta^*)}{\partial\beta'} & =\frac1N\sum_{i=1}^N\bigg\{ \frac{\partial}{\partial\beta'}\left[\frac{e(X_i'\beta^*)}{\Ex\{e(X'\beta^*)\}}\right]\frac{Z_iY_i}{e(X_i'\beta^*)} - \frac{\partial}{\partial\beta'}\left[\frac{e(X_i'\beta^*)h(X_i'\beta^*)}{\Ex\{e(X'\beta^*)h(X'\beta^*)\}}\right]\frac{(1-Z_i)Y_i}{1-e(X_i'\beta^*)}\\
    & \quad -\frac{f(X_i'\beta^*)e(X_i'\beta^*)}{\Ex\{e(X'\beta^*)\}}\frac{Z_iY_i}{e(X_i'\beta^*)^2}X_i - \frac{f(X_i'\beta^*)e(X_i'\beta^*)h(X_i'\beta^*)}{\Ex\{e(X'\beta^*)h(X'\beta^*)\}}\frac{(1-Z_i)Y_i}{\{1-e(X_i'\beta^*)\}^2}X_i
    \bigg\}\\
    & =\frac1N\sum_{i=1}^N\bigg\{ \frac{\partial}{\partial\beta'}\left[\frac{e(X_i'\beta^*)}{\Ex\{e(X'\beta^*)\}}\right]\frac{Z_iY_i}{e(X_i'\beta^*)} - \frac{\partial}{\partial\beta'}\left[\frac{e(X_i'\beta^*)h(X_i'\beta^*)}{\Ex\{e(X'\beta^*)h(X'\beta^*)\}}\right]\frac{(1-Z_i)Y_i}{1-e(X_i'\beta^*)}\\
    & \quad -\frac{f(X_i'\beta^*)}{\Ex\{e(X'\beta^*)\}}\frac{Z_iY_i}{e(X_i'\beta^*)}X_i - \frac{f(X_i'\beta^*)\omega_{0h}(X_i'\beta^*)}{\Ex\{e(X'\beta^*)h(X'\beta^*)\}}\frac{(1-Z_i)Y_i}{1-e(X_i'\beta^*)}X_i
    \bigg\}.
\end{align*}
Therefore, 
\begin{align}
    \Ex\left\{\dfrac{\partial\widehat\tau_{watt}^h(\mc H)(\beta^*)}{\partial\beta'}\right\} & 
    = \Ex\bigg\{\frac{\partial}{\partial\beta'}\left[\frac{e(X'\beta^*)}{\Ex\{e(X'\beta^*)\}}\right]\mu(1,X) -
    \frac{\partial}{\partial\beta'}\left[\frac{e(X'\beta^*)h(X'\beta^*)}{\Ex\{e(X'\beta^*)h(X'\beta^*)\}}\right]\mu(0,X) \label{eq:b1}\\
    & \quad -
    \left[\frac{\Ex\{X\mu(1,X)\mid e(X'\beta^*)\}}{\Ex\{e(X'\beta^*)\}} + \frac{\omega_{0h}(X'\beta^*)\Ex\{X\mu(0,X)\mid e(X'\beta^*)\}}{\Ex\{e(X'\beta^*)h(X'\beta^*)\}}\right]f(X'\beta^*)\bigg\}
    \label{eq:b2}\\
    & = b_1'-b_2',\nonumber
\end{align}
where $b_1'$ is \eqref{eq:b1}, and $b_2'$ is \eqref{eq:b2}. Therefore, 
\begin{align*}
    B' = (b_1-b_2)'\mc I(\beta^*)^{-1}. 
\end{align*}
Finally,
\begin{align}
    \sum_{k=1}^3\var(T_k) & = \Ex\left\{\mu\{1,e(X'\beta^*)\}^2\frac{e(X'\beta^*)}{\Ex\{e(X'\beta^*)\}^2} + \mu\{0,e(X'\beta^*)\}^2\frac{\omega_{0h}(X'\beta^*)^2\{1-e(X'\beta^*)\}}{\Ex\{e(X'\beta^*)h(X'\beta^*)\}^2}\right\}\label{eq:sigma1}\\
    & \quad + \Ex\left\{\sigma^2\{1,e(X'\beta^*)\}\frac{e(X'\beta^*)}{\Ex\{e(X'\beta^*)\}^2} + 
    \sigma^2\{0,e(X'\beta^*)\}\frac{\omega_{0h}(X'\beta^*)^2\{1-e(X'\beta^*)\}}{\Ex\{e(X'\beta^*)h(X'\beta^*)\}^2}\right\}\label{eq:sigma2}\\
    & \quad + \Ex\left\{\sigma^2(1,X)\frac{e(X'\beta^*)}{\Ex\{e(X'\beta^*)\}^2} + \sigma^2(0,X)\frac{\omega_{0h}(X'\beta^*)^2\{1-e(X'\beta^*)\}}{\Ex\{e(X'\beta^*)h(X'\beta^*)\}^2}\right\}\label{eq:sigma3}\\
    & \quad + 2B'\Ex\left\{\left[\frac{\Ex\{X\mu(1,X)\mid e(X'\beta^*)\}}{\Ex\{e(X'\beta^*)\}} + \frac{\omega_{0h}(X'\beta^*)\Ex\{X\mu(0,X)\mid e(X'\beta^*)\}}{\Ex\{e(X'\beta^*)h(X'\beta^*)\}}\right]f(X'\beta^*)\right\}\label{eq:bb1}\\
    & \quad + B'\mc I(\beta^*)B \label{eq:bb2}\\
    & = \sigma^2 + (b_1-b_2)'\mc I(\beta^*)^{-1}(b_1 + b_2)\nonumber \\
    & = \sigma^2 + b_1'\mc I(\beta^*)^{-1}b_1 - b_2'\mc I(\beta^*)^{-1}b_2,\nonumber
\end{align}
where the sum of \eqref{eq:sigma1}--\eqref{eq:sigma3} is $\sigma^2$, and the sum of \eqref{eq:bb1}--\eqref{eq:bb2} is $(b_1-b_2)'\mc I(\beta^*)^{-1}(b_1+b_2)=b_1'\mc I(\beta^*)^{-1}b_1 - b_2'\mc I(\beta^*)^{-1}b_2$. Note that \eqref{eq:bb1} $=2B'b_2$, and $b_1'\mc I(\beta^*)^{-1}b_2 = b_2'\mc I(\beta^*)^{-1}b_1$ because they are scalars. Thus the proof is completed. 

\subsection{Asymptotic behaviors under misspecified propensity score model}\label{subapx:asybeh-miss}

Consider an estimating equation $U_N(\theta) = \displaystyle\sum_{i=1}^N U_i(X_i,Z_i,Y_i;\theta) = 0$ for a parameter of interest, $\theta$. We solve the equation and get $\widehat\theta_N$ as the estimator of $\theta$ with truth $\theta^*$. Under regularity conditions, by Talyor expansion, we have
\begin{align*}
    0 = U_N(\widehat\theta_N) = U_N(\theta^*) + \frac{\partial U_N(\theta^*)}{\partial\theta'}(\widehat\theta_N-\theta^*) + o_p(1),
\end{align*}
therefore, the asymptotic bias of estimating $\theta^*$ using $\widehat\theta_N$ is 
\begin{align*}
    \widehat\theta_N-\theta^* = -\left\{\frac1N\sum_{i=1}^N\partial U_i(\theta^*)/\partial\theta'\right\}^{-1}\frac1NU_N(\theta^*) + o_p(1).
\end{align*}
Therefore, the asymptotic bias of $\widehat\theta_N$ is approximately
\begin{align}\label{eq:abias}
    -\Ex\{\partial U(X,Y,Z;\theta^*)/\partial\theta'\}^{-1}\Ex\{U(X,Y,Z;\theta^*)\}.
\end{align}

Denote the estimated PS by $\widehat e(X)$, which converges to some quantity $\widetilde e(X)$ such that $\widehat e(X)-\widetilde e(X)=o_p(1)$. If we correctly specify the PS model, then $\widetilde e(X)=e(X)+o_p(1)$, otherwise it may not be true. Let $\widehat\omega_{0h}(X) = \dfrac{\widehat e(X)\widehat h(X)}{1-\widehat e(X)}$ and $\widetilde\omega_{0h}(X)= \dfrac{\widetilde e(X)\widetilde h(X)}{1-\widetilde e(X)}$, where $\widehat h(X)$ (resp. $\widetilde h(X)$) is also by plugging-in $\widehat e(X)$ (resp. $\widetilde e(X)$). 

Now for $\widehat\tau_{watt}^h(\mc H)$, consider the following estimating equation,
\begin{align*}
    U_N(\mu_1,\mu_0) = \sum_{i=1}^N\left(Z_i(Y_i-\mu_1), ~~\widetilde\omega_{0h}(X_i)(1-Z_i)(Y_i-\mu_0)\right)' = 0,
\end{align*}
with the parameter of interest is $\theta = (\mu_1,\mu_0)'$, and $\widehat\tau_{watt}^h(\mc H) = c'\widehat\theta_N$ with $c=(1,-1)'$, where $\widehat\theta_N$ is by solving the equation. Therefore, the asymptotic bias of $\widehat\theta_N$, by \eqref{eq:abias}, is approximately
\begin{align*}
    \text{ABias}(\widehat\theta_N) & = \Ex
    \begin{pmatrix}
        Z & 0\\ 0 & \widetilde\omega_{0h}(X)(1-Z)
    \end{pmatrix}^{-1}
    \Ex
    \begin{pmatrix}
        Z(Y-\mu_1)\\ \widetilde\omega_{0h}(X)(1-Z)(Y-\mu_0)
    \end{pmatrix}\\
    & = 
    \begin{pmatrix}
       \Ex\{e(X)\}^{-1} & 0\\ 0 & \Ex\{\widetilde\omega_{0h}(X)\{1-e(X)\}\}^{-1}
    \end{pmatrix}
    \begin{pmatrix}
        0 \\ \Ex\{\widetilde\omega_{0h}(X)\{1-e(X)\}\Ex\{Y(0)-\mu_0\mid X\}\}
    \end{pmatrix}\\
    & = \frac{\Ex\{\widetilde\omega_{0h}(X)\{1-e(X)\}\Ex\{Y(0)-\mu_0\mid X\}\}}{\Ex\{\widetilde\omega_{0h}(X)\{1-e(X)\}\}}\\
    & = \frac{\Ex\{\widetilde\omega_{0h}(X)\{1-e(X)\}m_0(X)\}}{\Ex\{\widetilde\omega_{0h}(X)\{1-e(X)\}\}} - \mu_0,
\end{align*}
where $m_0(X)=\Ex\{Y(0)\mid X\}$. Hence,
\begin{align*}
    \text{ABias}(\widehat\tau_{watt}^h(\mc H)) & = \mu_0 - \frac{\Ex\{\widetilde\omega_{0h}(X)\{1-e(X)\}m_0(X)\}}{\Ex\{\widetilde\omega_{0h}(X)\{1-e(X)\}\}}\\
    & = \frac{\Ex\{\omega_{0h}(X)\{1-e(X)\}m_0(X)\}}{\Ex\{\omega_{0h}(X)\{1-e(X)\}\}} - \frac{\Ex\{\widetilde\omega_{0h}(X)\{1-e(X)\}m_0(X)\}}{\Ex\{\widetilde\omega_{0h}(X)\{1-e(X)\}\}}.
\end{align*}

\section{Appendix: Complete Simulation Results}\label{apx:simu}

\begin{table}[H]
\singlespacing
\centering
\scriptsize
\begin{tabular}{rccccccc}
  \toprule
  & \multicolumn{5}{c}{\color{black}Scenario}\\
  \cmidrule(lr){2-6}
  & \multicolumn{4}{c}{\color{black}DGP I} & \color{black}DGP II \\
  \cmidrule(lr){2-5}\cmidrule(lr){6-6}
Estimand & \color{black} \makecell[c]{Good \\ overlap} & \color{black} \makecell[c]{Moderate \\ overlap} & \color{black} \makecell[c]{Poor \\ overlap} &\color{black} \makecell[c]{Smaller \\ adjusted $R^2$} & \color{black}\makecell[c]{Kang and Schafer}\\ 
  \midrule
ATT & 23.750 & 18.312 & 15.960 & \color{black} 15.868 & \color{black} 20.000 \\ 
\addlinespace
  \bf OWATT & \bf 23.946 & \bf 18.894 & \bf 16.009 & \color{black} \bf 15.976 & \color{black} \bf 16.871 \\ 
  \addlinespace
  ATT trimming ($\alpha=0.05$) & 23.761 & 18.765 & 16.212 & \color{black} 16.166 & \color{black} 19.476 \\ 
  ATT trimming ($\alpha=0.10$) & 23.815 & 18.661 & 15.709 & \color{black} 15.677 & \color{black} 17.787\\ 
  ATT trimming ($\alpha=0.15$) & 23.900 & 18.376 & 15.134 & \color{black} 15.108 & \color{black} 15.583\\ 
  \addlinespace
  Smooth ATT trimming ($\alpha=0.05, \varepsilon=0.001$) & 23.761 & 18.765 & 16.212 & \color{black} 16.166 & \color{black} 19.475\\ 
  Smooth ATT trimming ($\alpha=0.10, \varepsilon=0.001$) & 23.815 & 18.661 & 15.710 & \color{black} 15.677 & \color{black} 17.790\\ 
  Smooth ATT trimming ($\alpha=0.15, \varepsilon=0.001$) & 23.900 & 18.376 & 15.134 & \color{black} 15.108 & \color{black} 15.582\\ 
  \addlinespace
  Smooth ATT trimming ($\alpha=0.05, \varepsilon=0.01$) & 23.753 & 18.652 & 16.365 & \color{black} 16.166 & \color{black} 19.452\\ 
  Smooth ATT trimming ($\alpha=0.10, \varepsilon=0.01$) & 23.769 & 18.692 & 16.176 & \color{black} 15.679 & \color{black} 17.780\\ 
  Smooth ATT trimming ($\alpha=0.15, \varepsilon=0.01$) & 23.802 & 18.637 & 15.940 & \color{black} 15.111 & \color{black} 15.582\\
  \addlinespace
  Smooth ATT trimming ($\alpha=0.05, \varepsilon=0.05$) & 23.762 & 18.758 & 16.212 & \color{black} 16.037 & \color{black} 19.030\\ 
  Smooth ATT trimming ($\alpha=0.10, \varepsilon=0.05$) & 23.815 & 18.658 & 15.711 & \color{black} 15.694 & \color{black} 17.614\\ 
  Smooth ATT trimming ($\alpha=0.15, \varepsilon=0.05$) & 23.900 & 18.374 & 15.137 & \color{black} 15.170 & \color{black} 15.643\\
  \addlinespace
  ATT truncation ($\alpha=0.05$) & 23.779 & 18.607 & 16.052 & \color{black} 16.309 & \color{black} 19.850\\ 
  ATT truncation ($\alpha=0.10$) & 23.827 & 18.574 & 15.725 & \color{black} 16.128 & \color{black} 19.245\\ 
  ATT truncation ($\alpha=0.15$) & 23.890 & 18.334 & 15.195 & \color{black} 15.898 & \color{black} 18.342\\ 
   \bottomrule
\end{tabular}
\caption{True values of causal effects on the treated, using $10$ independently generated super-population of $10^6$ size under the true data generating process and true propensity scores, and taking average}\label{tab:truth}
\end{table}

\textcolor{black}{\subsection{Complete simulation results by DGP I}\label{subapp:add-ours}}

\begin{figure}[H]
    \centering
    \includegraphics[trim=10 05 10 05, clip,  width=0.8\textwidth]{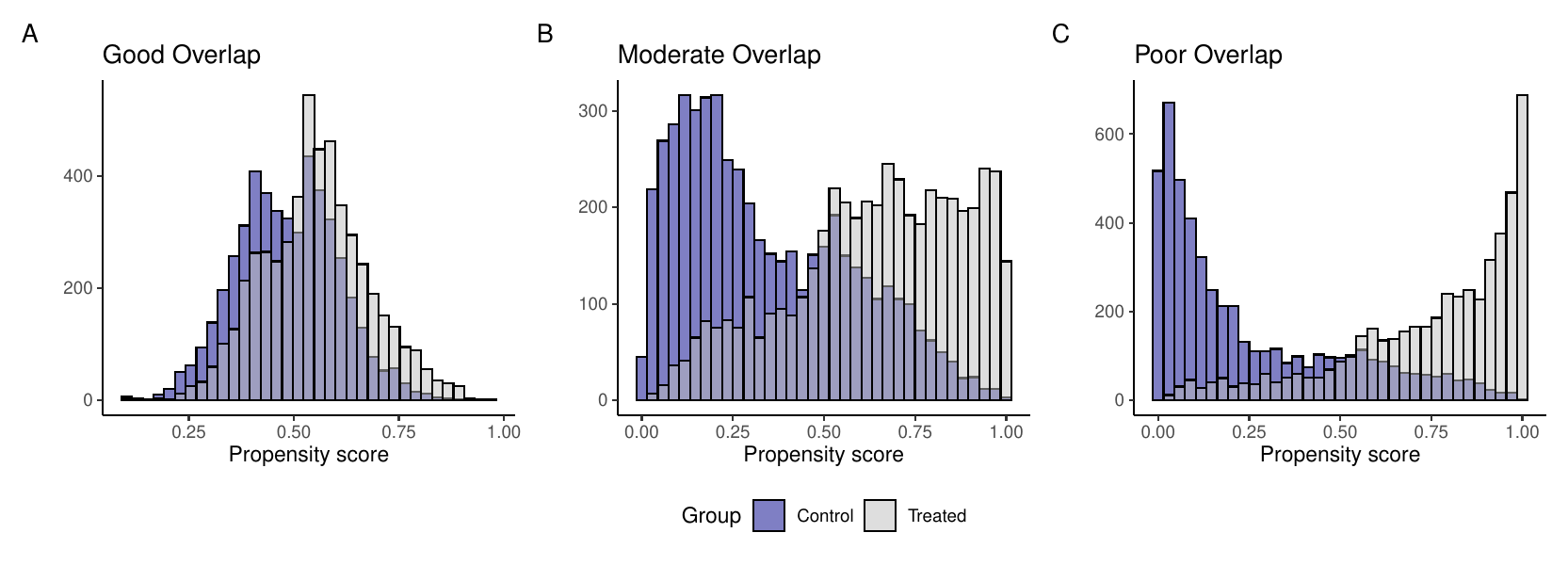}
    \caption{Propensity score distributions by treatment group of 3 overlap levels in simulation under DGP I, using a randomly generated sample of $N=10^4$.}
    \label{fig:ps-distns}
\end{figure}
% \begin{figure}[H]
%     \centering
%     \includegraphics[trim=05 05 10 05, clip, width=0.9\textwidth]{rbias_good.pdf}
%     \caption{Boxplots of relative percent bias under good overlap over 1000 Monte Carlo replications}
%     \label{fig:rbias_good}
% \end{figure}

% \begin{figure}[H]
%     \centering
%     \includegraphics[trim=05 05 10 05, clip, width=0.9\textwidth]{rbias_mode.pdf}
%     \caption{Boxplots of relative percent bias under moderate overlap over 1000 Monte Carlo replications}
%     \label{fig:rbias_mode}
% \end{figure}

% \begin{figure}[H]
%     \centering
%     \includegraphics[trim=05 05 10 05, clip, width=0.9\textwidth]{rbias_poor.pdf}
%     \begin{tablenotes}\footnotesize
%         \item For smooth ATT trimming methods, the parameter in the bracket is $(\alpha,\varepsilon)$, i.e., trimming threshold and standard error of the normal cdf in the tilting function, respectively. 
%     \end{tablenotes}
%     \caption{Boxplots of relative percent bias under poor overlap over 1000 Monte Carlo replications}
%     \label{fig:rbias_poor}
% \end{figure}

\begin{table}[H]
    \singlespacing
    \scriptsize
    \centering
    \begin{tabular}{rccccc}
    \toprule
        Method & ARBias\% & RRMSE & RMSE & RE & CP\% \\
        \midrule
         & \multicolumn{5}{c}{PS model correctly specified} \\
         \cmidrule(rl){2-6}
        ATT & 0.29 & 0.037 & 0.87 & 0.96 & 95.20 \\ 
        \addlinespace
        \bf OWATT &\bf  0.22 &\bf  0.035 & \bf 0.83 & \bf 0.95 &\bf 95.70 \\ 
        \addlinespace
        ATT trimming ($\alpha=0.05$), PS re-estimated & 0.31 & 0.037 & 0.87 & 0.96 & 95.40 \\ 
  ATT trimming ($\alpha=0.10$), PS re-estimated & 0.27 & 0.037 & 0.87 & 0.96 & 95.40 \\ 
  ATT trimming ($\alpha=0.15$), PS re-estimated & 0.21 & 0.037 & 0.87 & 0.96 & 95.40 \\ 
        \addlinespace
     ATT trimming ($\alpha=0.05$), PS not re-estimated & 0.31 & 0.037 & 0.87 & 0.96 & 95.40 \\ 
  ATT trimming ($\alpha=0.10$), PS not re-estimated & 0.31 & 0.037 & 0.88 & 0.96 & 95.50 \\ 
  ATT trimming ($\alpha=0.15$), PS not re-estimated & 0.26 & 0.037 & 0.88 & 0.97 & 95.30 \\ 
         \addlinespace
        Smooth ATT trimming ($\alpha=0.05, \varepsilon=0.001$) & 0.31 & 0.037 & 0.87 & 0.96 & 95.40 \\ 
  Smooth ATT trimming ($\alpha=0.10, \varepsilon=0.001$) & 0.31 & 0.037 & 0.88 & 0.96 & 95.40 \\ 
  Smooth ATT trimming ($\alpha=0.15, \varepsilon=0.001$) & 0.26 & 0.037 & 0.88 & 0.97 & 95.40 \\ 
        \addlinespace
        Smooth ATT trimming ($\alpha=0.05, \varepsilon=0.01$) & 0.31 & 0.037 & 0.87 & 0.96 & 95.40 \\ 
  Smooth ATT trimming ($\alpha=0.10, \varepsilon=0.01$) & 0.31 & 0.037 & 0.87 & 0.96 & 95.30 \\ 
  Smooth ATT trimming ($\alpha=0.15, \varepsilon=0.01$) & 0.26 & 0.037 & 0.88 & 0.97 & 95.40 \\ 
        \addlinespace
        Smooth ATT trimming ($\alpha=0.05, \varepsilon=0.05$) & 0.30 & 0.037 & 0.87 & 0.96 & 95.60 \\ 
  Smooth ATT trimming ($\alpha=0.10, \varepsilon=0.05$) & 0.29 & 0.036 & 0.87 & 0.96 & 95.30 \\ 
  Smooth ATT trimming ($\alpha=0.15, \varepsilon=0.05$) & 0.25 & 0.036 & 0.87 & 0.96 & 95.60 \\  
        \addlinespace
        ATT truncation ($\alpha=0.05$) & 0.30 & 0.037 & 0.87 & 0.97 & 95.20 \\ 
  ATT truncation ($\alpha=0.10$) & 0.31 & 0.037 & 0.87 & 0.96 & 95.50 \\ 
  ATT truncation ($\alpha=0.15$) & 0.30 & 0.037 & 0.87 & 0.96 & 95.50 \\ 
        \addlinespace
        \midrule & \multicolumn{5}{c}{PS model misspecified}\\
         \cmidrule(rl){2-6}
        ATT & 2.52 & 0.051 & 1.20 & 0.99 & 92.60 \\ 
        \addlinespace
       \bf OWATT &  \bf 2.46 &  \bf 0.049 &  \bf 1.17 &  \bf 0.98 &  \bf 92.70 \\ 
        \addlinespace
        ATT trimming ($\alpha=0.05$), PS re-estimated & 2.48 & 0.050 & 1.19 & 0.99 & 92.70 \\ 
  ATT trimming ($\alpha=0.10$), PS re-estimated & 2.25 & 0.049 & 1.17 & 0.99 & 93.30 \\ 
  ATT trimming ($\alpha=0.15$), PS re-estimated & 1.97 & 0.047 & 1.13 & 0.98 & 93.80 \\ 
        \addlinespace
     ATT trimming ($\alpha=0.05$), PS not re-estimated & 2.48 & 0.050 & 1.19 & 0.99 & 92.70 \\ 
  ATT trimming ($\alpha=0.10$), PS not re-estimated & 2.25 & 0.049 & 1.17 & 0.99 & 93.30 \\ 
  ATT trimming ($\alpha=0.15$), PS not re-estimated & 1.95 & 0.047 & 1.13 & 0.98 & 93.90 \\ 
      \addlinespace
        Smooth ATT trimming ($\alpha=0.05, \varepsilon=0.001$) & 2.48 & 0.050 & 1.19 & 0.99 & 92.70 \\ 
  Smooth ATT trimming ($\alpha=0.10, \varepsilon=0.001$) & 2.25 & 0.049 & 1.17 & 0.99 & 93.30 \\ 
  Smooth ATT trimming ($\alpha=0.15, \varepsilon=0.001$) & 1.95 & 0.047 & 1.13 & 0.98 & 93.90 \\ 
        \addlinespace
        Smooth ATT trimming ($\alpha=0.05, \varepsilon=0.01$) & 2.47 & 0.050 & 1.19 & 0.99 & 92.70 \\ 
  Smooth ATT trimming ($\alpha=0.10, \varepsilon=0.01$) & 2.25 & 0.049 & 1.17 & 0.99 & 93.30 \\ 
  Smooth ATT trimming ($\alpha=0.15, \varepsilon=0.01$) & 1.95 & 0.047 & 1.13 & 0.98 & 93.70 \\ 
        \addlinespace
        Smooth ATT trimming ($\alpha=0.05, \varepsilon=0.05$) & 2.41 & 0.050 & 1.18 & 0.99 & 92.90 \\ 
  Smooth ATT trimming ($\alpha=0.10, \varepsilon=0.05$) & 2.23 & 0.049 & 1.16 & 0.98 & 93.40 \\ 
  Smooth ATT trimming ($\alpha=0.15, \varepsilon=0.05$) & 2.06 & 0.048 & 1.14 & 0.98 & 93.80 \\ 
        \addlinespace
       ATT truncation ($\alpha=0.05$) & 2.51 & 0.050 & 1.20 & 0.99 & 92.60 \\ 
  ATT truncation ($\alpha=0.10$) & 2.44 & 0.050 & 1.19 & 0.99 & 92.70 \\ 
  ATT truncation ($\alpha=0.15$) & 2.31 & 0.049 & 1.17 & 0.99 & 93.00 \\
       \bottomrule
    \end{tabular}
    \begin{tablenotes}\scriptsize
        \item ATT: average treatment effect on the treated; OWATT: overlap weighted average treatment effect on the treated; PS: propensity score; ARBias\%: absolute relative percent bias; RRMSE: relative root mean square error; RMSE: root mean square error; RE: relative efficiency; CP\%: coverage probability (\%). 
    \end{tablenotes}
    \caption{Full simulation results of good overlap PS model}\label{tab:res_good}
\end{table}

\begin{table}[H]
\singlespacing
    \scriptsize
    \centering
    \begin{tabular}{rccccc}
    \toprule
        Method & ARBias\% & RRMSE & RMSE & RE & CP\% \\
        \midrule
         & \multicolumn{5}{c}{PS model correctly specified} \\
         \cmidrule(rl){2-6}
        ATT & 0.50 & 0.056 & 1.02 & 1.18 & 94.40 \\ 
        \addlinespace
 \bf OWATT & \bf 0.15 & \bf 0.039 & \bf 0.73 & \bf 0.99 & \bf 94.80 \\ 
        \addlinespace
        ATT trimming ($\alpha=0.05$), PS re-estimated & 0.02 & 0.040 & 0.75 & 0.98 & 95.20 \\ 
  ATT trimming ($\alpha=0.10$), PS re-estimated & 0.55 & 0.041 & 0.76 & 1.00 & 95.00 \\ 
  ATT trimming ($\alpha=0.15$), PS re-estimated & 1.39 & 0.043 & 0.79 & 0.99 & 94.20 \\ 
        \addlinespace
     ATT trimming ($\alpha=0.05$), PS not re-estimated & 0.16 & 0.040 & 0.75 & 0.98 & 95.00 \\ 
  ATT trimming ($\alpha=0.10$), PS not re-estimated & 0.19 & 0.040 & 0.74 & 1.00 & 94.90 \\ 
  ATT trimming ($\alpha=0.15$), PS not re-estimated & 0.18 & 0.040 & 0.74 & 0.99 & 94.20 \\ 
         \addlinespace
        Smooth ATT trimming ($\alpha=0.05, \varepsilon=0.001$) & 0.16 & 0.040 & 0.75 & 0.98 & 95.00 \\ 
  Smooth ATT trimming ($\alpha=0.10, \varepsilon=0.001$) & 0.19 & 0.040 & 0.74 & 1.00 & 94.80 \\ 
  Smooth ATT trimming ($\alpha=0.15, \varepsilon=0.001$) & 0.18 & 0.040 & 0.74 & 0.99 & 94.20 \\ 
        \addlinespace
        Smooth ATT trimming ($\alpha=0.05, \varepsilon=0.01$) & 0.16 & 0.040 & 0.75 & 0.99 & 94.90 \\ 
  Smooth ATT trimming ($\alpha=0.10, \varepsilon=0.01$) & 0.18 & 0.040 & 0.74 & 1.00 & 94.70 \\ 
  Smooth ATT trimming ($\alpha=0.15, \varepsilon=0.01$) & 0.18 & 0.040 & 0.74 & 1.00 & 94.30 \\ 
        \addlinespace
        Smooth ATT trimming ($\alpha=0.05, \varepsilon=0.05$) & 0.06 & 0.041 & 0.76 & 0.98 & 95.40 \\ 
  Smooth ATT trimming ($\alpha=0.10, \varepsilon=0.05$) & 0.15 & 0.040 & 0.74 & 0.99 & 94.60 \\ 
  Smooth ATT trimming ($\alpha=0.15, \varepsilon=0.05$) & 0.18 & 0.040 & 0.73 & 1.00 & 94.50 \\ 
        \addlinespace
ATT truncation ($\alpha=0.05$) & 0.10 & 0.041 & 0.77 & 1.00 & 95.20 \\ 
  ATT truncation ($\alpha=0.10$) & 0.14 & 0.040 & 0.75 & 1.00 & 95.10 \\ 
  ATT truncation ($\alpha=0.15$) & 0.15 & 0.040 & 0.74 & 1.00 & 95.00 \\
        \addlinespace
        \midrule & \multicolumn{5}{c}{PS model misspecified}\\
         \cmidrule(rl){2-6}
        ATT & 3.93 & 0.074 & 1.35 & 1.16 & 83.40 \\ 
        \addlinespace
        \bf OWATT & \bf  4.23 &  \bf 0.061 &  \bf 1.16 &  \bf 1.01 &  \bf 85.30 \\ 
        \addlinespace
        ATT trimming ($\alpha=0.05$), PS re-estimated & 3.88 & 0.059 & 1.11 & 1.00 & 86.60 \\ 
  ATT trimming ($\alpha=0.10$), PS re-estimated & 5.37 & 0.069 & 1.29 & 1.00 & 78.80 \\ 
  ATT trimming ($\alpha=0.15$), PS re-estimated & 6.90 & 0.082 & 1.51 & 1.00 & 66.80 \\ 
        \addlinespace
     ATT trimming ($\alpha=0.05$), PS not re-estimated & 3.53 & 0.057 & 1.07 & 1.00 & 88.50 \\ 
  ATT trimming ($\alpha=0.10$), PS not re-estimated & 3.94 & 0.059 & 1.11 & 1.00 & 86.70 \\ 
  ATT trimming ($\alpha=0.15$), PS not re-estimated & 4.03 & 0.061 & 1.12 & 1.00 & 86.60 \\ 
       \addlinespace
        Smooth ATT trimming ($\alpha=0.05, \varepsilon=0.001$) & 3.53 & 0.057 & 1.07 & 1.00 & 88.50 \\ 
  Smooth ATT trimming ($\alpha=0.10, \varepsilon=0.001$) & 3.94 & 0.059 & 1.11 & 1.00 & 86.80 \\ 
  Smooth ATT trimming ($\alpha=0.15, \varepsilon=0.001$) & 4.03 & 0.061 & 1.12 & 1.00 & 86.70 \\ 
        \addlinespace
        Smooth ATT trimming ($\alpha=0.05, \varepsilon=0.01$) & 3.52 & 0.057 & 1.07 & 1.01 & 88.60 \\ 
  Smooth ATT trimming ($\alpha=0.10, \varepsilon=0.01$) & 3.93 & 0.059 & 1.11 & 1.00 & 86.70 \\ 
  Smooth ATT trimming ($\alpha=0.15, \varepsilon=0.01$) & 4.03 & 0.061 & 1.12 & 1.00 & 86.60 \\  
        \addlinespace
        Smooth ATT trimming ($\alpha=0.05, \varepsilon=0.05$) & 3.63 & 0.059 & 1.11 & 1.03 & 87.40 \\ 
  Smooth ATT trimming ($\alpha=0.10, \varepsilon=0.05$) & 3.82 & 0.059 & 1.10 & 1.01 & 87.20 \\ 
  Smooth ATT trimming ($\alpha=0.15, \varepsilon=0.05$) & 4.01 & 0.061 & 1.11 & 1.00 & 86.80 \\ 
        \addlinespace
        ATT truncation ($\alpha=0.05$) & 3.22 & 0.058 & 1.07 & 1.03 & 88.90 \\ 
  ATT truncation ($\alpha=0.10$) & 3.50 & 0.058 & 1.08 & 1.02 & 87.70 \\ 
  ATT truncation ($\alpha=0.15$) & 3.67 & 0.058 & 1.08 & 1.01 & 87.40 \\ 
       \bottomrule
    \end{tabular}
    \begin{tablenotes}\scriptsize
        \item ATT: average treatment effect on the treated; OWATT: overlap weighted average treatment effect on the treated; PS: propensity score; ARBias\%: absolute relative percent bias; RRMSE: relative root mean square error; RMSE: root mean square error; RE: relative efficiency; CP\%: coverage probability (\%). 
    \end{tablenotes}
    \caption{Full simulation results of moderate overlap PS model}\label{tab:res_mode}
\end{table}

\begin{table}[H]
    \singlespacing
    \scriptsize
    \centering
    \begin{tabular}{rccccc}
    \toprule
        Method & ARBias\% & RRMSE & RMSE & RE & CP\% \\
        \midrule
         & \multicolumn{5}{c}{PS model correctly specified} \\
         \cmidrule(rl){2-6}
        ATT & 1.64 & 0.074 & 1.19 & 1.51 & 91.20 \\ 
        \addlinespace
  \bf OWATT & \bf 0.13 & \bf 0.041 &\bf 0.66 & \bf 0.98 & \bf 94.40 \\ 
        \addlinespace
        ATT trimming ($\alpha=0.05$), PS re-estimated & 0.68 & 0.043 & 0.70 & 0.97 & 94.90 \\ 
  ATT trimming ($\alpha=0.10$), PS re-estimated & 1.67 & 0.047 & 0.73 & 0.95 & 94.50 \\ 
  ATT trimming ($\alpha=0.15$), PS re-estimated & 2.81 & 0.054 & 0.82 & 0.96 & 91.20 \\
        \addlinespace
     ATT trimming ($\alpha=0.05$), PS not re-estimated & 0.08 & 0.042 & 0.68 & 0.97 & 94.80 \\ 
  ATT trimming ($\alpha=0.10$), PS not re-estimated & 0.18 & 0.043 & 0.68 & 0.95 & 94.90 \\ 
  ATT trimming ($\alpha=0.15$), PS not re-estimated & 0.21 & 0.046 & 0.69 & 0.99 & 94.10 \\
        \addlinespace
        Smooth ATT trimming ($\alpha=0.05, \varepsilon=0.001$) & 0.09 & 0.042 & 0.68 & 0.97 & 95.00 \\ 
  Smooth ATT trimming ($\alpha=0.10, \varepsilon=0.001$) & 0.18 & 0.043 & 0.68 & 0.96 & 94.90 \\ 
  Smooth ATT trimming ($\alpha=0.15, \varepsilon=0.001$) & 0.21 & 0.046 & 0.69 & 0.99 & 94.20 \\
        \addlinespace
        Smooth ATT trimming ($\alpha=0.05, \varepsilon=0.01$) & 0.10 & 0.042 & 0.68 & 0.98 & 94.90 \\ 
  Smooth ATT trimming ($\alpha=0.10, \varepsilon=0.01$) & 0.18 & 0.043 & 0.68 & 0.97 & 94.70 \\ 
  Smooth ATT trimming ($\alpha=0.15, \varepsilon=0.01$) & 0.20 & 0.045 & 0.69 & 0.99 & 94.40 \\
        \addlinespace
       Smooth ATT trimming ($\alpha=0.05, \varepsilon=0.05$) & 0.13 & 0.046 & 0.74 & 1.05 & 94.70 \\ 
  Smooth ATT trimming ($\alpha=0.10, \varepsilon=0.05$) & 0.11 & 0.042 & 0.66 & 0.98 & 94.40 \\ 
  Smooth ATT trimming ($\alpha=0.15, \varepsilon=0.05$) & 0.17 & 0.044 & 0.66 & 0.98 & 94.20 \\ 
        \addlinespace
        ATT truncation ($\alpha=0.05$) & 0.11 & 0.043 & 0.70 & 1.00 & 94.30 \\ 
  ATT truncation ($\alpha=0.10$) & 0.12 & 0.042 & 0.68 & 0.99 & 94.20 \\ 
  ATT truncation ($\alpha=0.15$) & 0.13 & 0.042 & 0.67 & 0.99 & 93.90 \\ 
        \addlinespace
        \midrule & \multicolumn{5}{c}{PS model misspecified}\\
         \cmidrule(rl){2-6}
        ATT & 2.32 & 0.184 & 2.93 & 2.42 & 69.20 \\ 
        \addlinespace
  \bf OWATT & \bf 5.91 & \bf 0.075 & \bf 1.20 & \bf 1.00 & \bf 76.80 \\ 
        \addlinespace
        ATT trimming ($\alpha=0.05$), PS re-estimated & 6.60 & 0.080 & 1.30 & 0.99 & 70.80 \\ 
  ATT trimming ($\alpha=0.10$), PS re-estimated & 7.78 & 0.090 & 1.42 & 0.98 & 62.80 \\ 
  ATT trimming ($\alpha=0.15$), PS re-estimated & 9.45 & 0.106 & 1.60 & 0.99 & 49.90 \\ 
        \addlinespace
     ATT trimming ($\alpha=0.05$), PS not re-estimated & 4.80 & 0.066 & 1.07 & 0.98 & 84.30 \\ 
  ATT trimming ($\alpha=0.10$), PS not re-estimated & 4.56 & 0.065 & 1.02 & 0.97 & 86.10 \\ 
  ATT trimming ($\alpha=0.15$), PS not re-estimated & 4.95 & 0.069 & 1.05 & 0.98 & 84.10 \\ 
       \addlinespace
        Smooth ATT trimming ($\alpha=0.05, \varepsilon=0.001$) & 4.80 & 0.066 & 1.07 & 0.99 & 84.10 \\ 
  Smooth ATT trimming ($\alpha=0.10, \varepsilon=0.001$) & 4.55 & 0.065 & 1.02 & 0.97 & 86.10 \\ 
  Smooth ATT trimming ($\alpha=0.15, \varepsilon=0.001$) & 4.95 & 0.069 & 1.05 & 0.98 & 84.10 \\ 
        \addlinespace
        Smooth ATT trimming ($\alpha=0.05, \varepsilon=0.01$) & 4.84 & 0.066 & 1.07 & 0.99 & 83.30 \\ 
  Smooth ATT trimming ($\alpha=0.10, \varepsilon=0.01$) & 4.57 & 0.065 & 1.02 & 0.97 & 85.80 \\ 
  Smooth ATT trimming ($\alpha=0.15, \varepsilon=0.01$) & 4.97 & 0.069 & 1.05 & 0.98 & 83.90 \\ 
        \addlinespace
        Smooth ATT trimming ($\alpha=0.05, \varepsilon=0.05$) & 3.67 & 0.098 & 1.58 & 1.62 & 80.60 \\ 
  Smooth ATT trimming ($\alpha=0.10, \varepsilon=0.05$) & 4.63 & 0.069 & 1.08 & 1.04 & 82.80 \\ 
  Smooth ATT trimming ($\alpha=0.15, \varepsilon=0.05$) & 5.11 & 0.070 & 1.06 & 0.99 & 83.20 \\ 
        \addlinespace
        ATT truncation ($\alpha=0.05$) & 5.09 & 0.069 & 1.14 & 1.02 & 80.70 \\ 
  ATT truncation ($\alpha=0.10$) & 4.95 & 0.067 & 1.09 & 1.01 & 82.30 \\ 
  ATT truncation ($\alpha=0.15$) & 4.96 & 0.067 & 1.07 & 1.00 & 82.80 \\ 
       \bottomrule
    \end{tabular}
    \begin{tablenotes}\scriptsize
        \item ATT: average treatment effect on the treated; OWATT: overlap weighted average treatment effect on the treated; PS: propensity score; ARBias\%: absolute relative percent bias; RRMSE: relative root mean square error; RMSE: root mean square error; RE: relative efficiency; CP\%: coverage probability (\%). 
    \end{tablenotes}
    \caption{Full simulation results of poor overlap PS model}\label{tab:res_poor}
\end{table}

\begin{table}[H]
    \singlespacing
    \scriptsize
    \centering
    \textcolor{black}{
    \begin{tabular}{rccccc}
    \toprule
        Method & ARBias\% & RRMSE & RMSE & RE & CP\% \\
        \midrule
         & \multicolumn{5}{c}{PS model correctly specified} \\
         \cmidrule(rl){2-6}
        ATT & 2.47 & 0.128 & 2.03 & 1.91 & 93.20 \\ 
        \addlinespace
       \bf OWATT & \bf 0.14 & \bf 0.062 & \bf 1.00 & \bf 0.94 & \bf 96.00 \\ 
        \addlinespace
        ATT trimming ($\alpha=0.05$), PS re-estimated & 0.90 & 0.075 & 1.21 & 0.94 & 96.70 \\ 
  ATT trimming ($\alpha=0.10$), PS re-estimated & 1.89 & 0.073 & 1.15 & 0.89 & 95.50 \\ 
  ATT trimming ($\alpha=0.15$), PS re-estimated & 2.98 & 0.081 & 1.22 & 0.94 & 95.00 \\ 
        \addlinespace
     ATT trimming ($\alpha=0.05$), PS not re-estimated & 0.13 & 0.071 & 1.15 & 0.93 & 96.40 \\ 
  ATT trimming ($\alpha=0.10$), PS not re-estimated & 0.06 & 0.067 & 1.06 & 0.90 & 97.20 \\ 
  ATT trimming ($\alpha=0.15$), PS not re-estimated & 0.03 & 0.069 & 1.05 & 0.94 & 95.60 \\ 
         \addlinespace
        Smooth ATT trimming ($\alpha=0.05, \varepsilon=0.001$) & 0.14 & 0.071 & 1.14 & 0.92 & 96.40 \\ 
  Smooth ATT trimming ($\alpha=0.10, \varepsilon=0.001$) & 0.05 & 0.067 & 1.06 & 0.90 & 97.30 \\ 
  Smooth ATT trimming ($\alpha=0.15, \varepsilon=0.001$) & 0.04 & 0.069 & 1.05 & 0.94 & 95.60 \\ 
        \addlinespace
        Smooth ATT trimming ($\alpha=0.05, \varepsilon=0.01$) & 0.15 & 0.069 & 1.12 & 0.93 & 96.40 \\ 
  Smooth ATT trimming ($\alpha=0.10, \varepsilon=0.01$) & 0.08 & 0.066 & 1.03 & 0.90 & 97.00 \\ 
  Smooth ATT trimming ($\alpha=0.15, \varepsilon=0.01$) & 0.03 & 0.068 & 1.03 & 0.94 & 95.70 \\ 
        \addlinespace
        Smooth ATT trimming ($\alpha=0.05, \varepsilon=0.05$) & 0.36 & 0.072 & 1.15 & 1.01 & 96.40 \\ 
  Smooth ATT trimming ($\alpha=0.10, \varepsilon=0.05$) & 0.15 & 0.064 & 1.00 & 0.92 & 96.70 \\ 
  Smooth ATT trimming ($\alpha=0.15, \varepsilon=0.05$) & 0.04 & 0.064 & 0.98 & 0.94 & 95.50 \\ 
        \addlinespace
        ATT truncation ($\alpha=0.05$) & 0.34 & 0.074 & 1.21 & 0.98 & 96.20 \\ 
  ATT truncation ($\alpha=0.10$) & 0.23 & 0.067 & 1.08 & 0.93 & 96.90 \\ 
  ATT truncation ($\alpha=0.15$) & 0.20 & 0.065 & 1.03 & 0.92 & 96.50 \\ 
        \addlinespace
        \midrule & \multicolumn{5}{c}{PS model misspecified}\\
         \cmidrule(rl){2-6}
        ATT & 3.45 & 0.198 & 3.14 & 3.68 & 85.80 \\ 
        \addlinespace
        \bf OWATT & \bf 4.71 & \bf 0.080 & \bf 1.27 & \bf 0.94 & \bf 90.00 \\ 
        \addlinespace
        ATT trimming ($\alpha=0.05$), PS re-estimated & 5.41 & 0.092 & 1.49 & 0.91 & 91.30 \\ 
  ATT trimming ($\alpha=0.10$), PS re-estimated & 7.01 & 0.102 & 1.60 & 0.95 & 85.40 \\ 
  ATT trimming ($\alpha=0.15$), PS re-estimated & 8.89 & 0.117 & 1.76 & 0.94 & 80.40 \\ 
        \addlinespace
     ATT trimming ($\alpha=0.05$), PS not re-estimated & 3.91 & 0.082 & 1.32 & 0.91 & 93.30 \\ 
  ATT trimming ($\alpha=0.10$), PS not re-estimated & 4.08 & 0.082 & 1.28 & 0.95 & 91.90 \\ 
  ATT trimming ($\alpha=0.15$), PS not re-estimated & 4.58 & 0.085 & 1.28 & 0.96 & 90.70 \\ 
      \addlinespace
        Smooth ATT trimming ($\alpha=0.05, \varepsilon=0.001$) & 3.91 & 0.081 & 1.32 & 0.91 & 93.50 \\ 
  Smooth ATT trimming ($\alpha=0.10, \varepsilon=0.001$) & 4.09 & 0.081 & 1.28 & 0.95 & 91.90 \\ 
  Smooth ATT trimming ($\alpha=0.15, \varepsilon=0.001$) & 4.58 & 0.085 & 1.28 & 0.96 & 90.90 \\ 
        \addlinespace
        Smooth ATT trimming ($\alpha=0.05, \varepsilon=0.01$) & 3.97 & 0.080 & 1.30 & 0.92 & 93.30 \\ 
  Smooth ATT trimming ($\alpha=0.10, \varepsilon=0.01$) & 4.06 & 0.080 & 1.26 & 0.95 & 91.50 \\ 
  Smooth ATT trimming ($\alpha=0.15, \varepsilon=0.01$) & 4.55 & 0.083 & 1.26 & 0.95 & 90.70 \\ 
        \addlinespace
        Smooth ATT trimming ($\alpha=0.05, \varepsilon=0.05$) & 3.35 & 0.100 & 1.61 & 1.56 & 90.80 \\ 
  Smooth ATT trimming ($\alpha=0.10, \varepsilon=0.05$) & 4.09 & 0.079 & 1.24 & 0.98 & 92.00 \\ 
  Smooth ATT trimming ($\alpha=0.15, \varepsilon=0.05$) & 4.60 & 0.081 & 1.23 & 0.95 & 89.90 \\ 
        \addlinespace
       ATT truncation ($\alpha=0.05$) & 4.00 & 0.087 & 1.41 & 0.98 & 92.20 \\ 
  ATT truncation ($\alpha=0.10$) & 4.04 & 0.080 & 1.29 & 0.95 & 91.80 \\ 
  ATT truncation ($\alpha=0.15$) & 4.17 & 0.079 & 1.25 & 0.95 & 91.90 \\ 
       \bottomrule
    \end{tabular}}
    \begin{tablenotes}\scriptsize
        \item ATT: average treatment effect on the treated; OWATT: overlap weighted average treatment effect on the treated; PS: propensity score; ARBias\%: absolute relative percent bias; RRMSE: relative root mean square error; RMSE: root mean square error; RE: relative efficiency; CP\%: coverage probability (\%). 
    \end{tablenotes}
    \caption{Full simulation results of poor overlap PS model with a smaller adjusted $R^2$}\label{tab:res_poor-smallR2}
\end{table}

\begin{figure}[H]
    \centering
    \includegraphics[width=\textwidth]{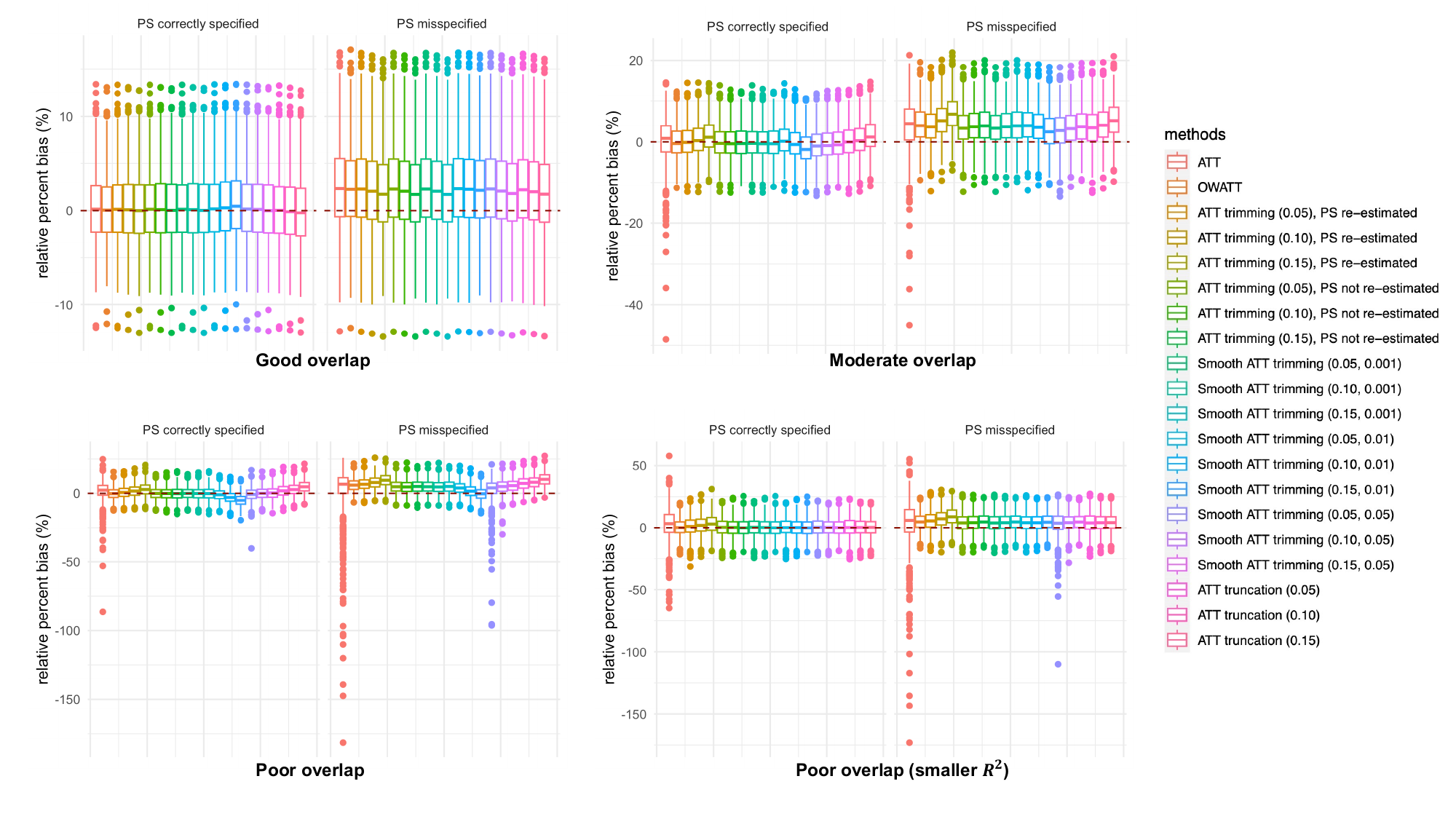}
    \caption{\color{black}Boxplots of relative percent biases by cases under DGP I}
    \label{fig:Rbias-B.1}
\end{figure}

\textcolor{black}{\subsection{Complete simulation results by DGP II}\label{subapp:add-Kang}}

\begin{figure}[H]
    \centering
    \includegraphics[trim=10 100 10 100, clip, width=\textwidth]{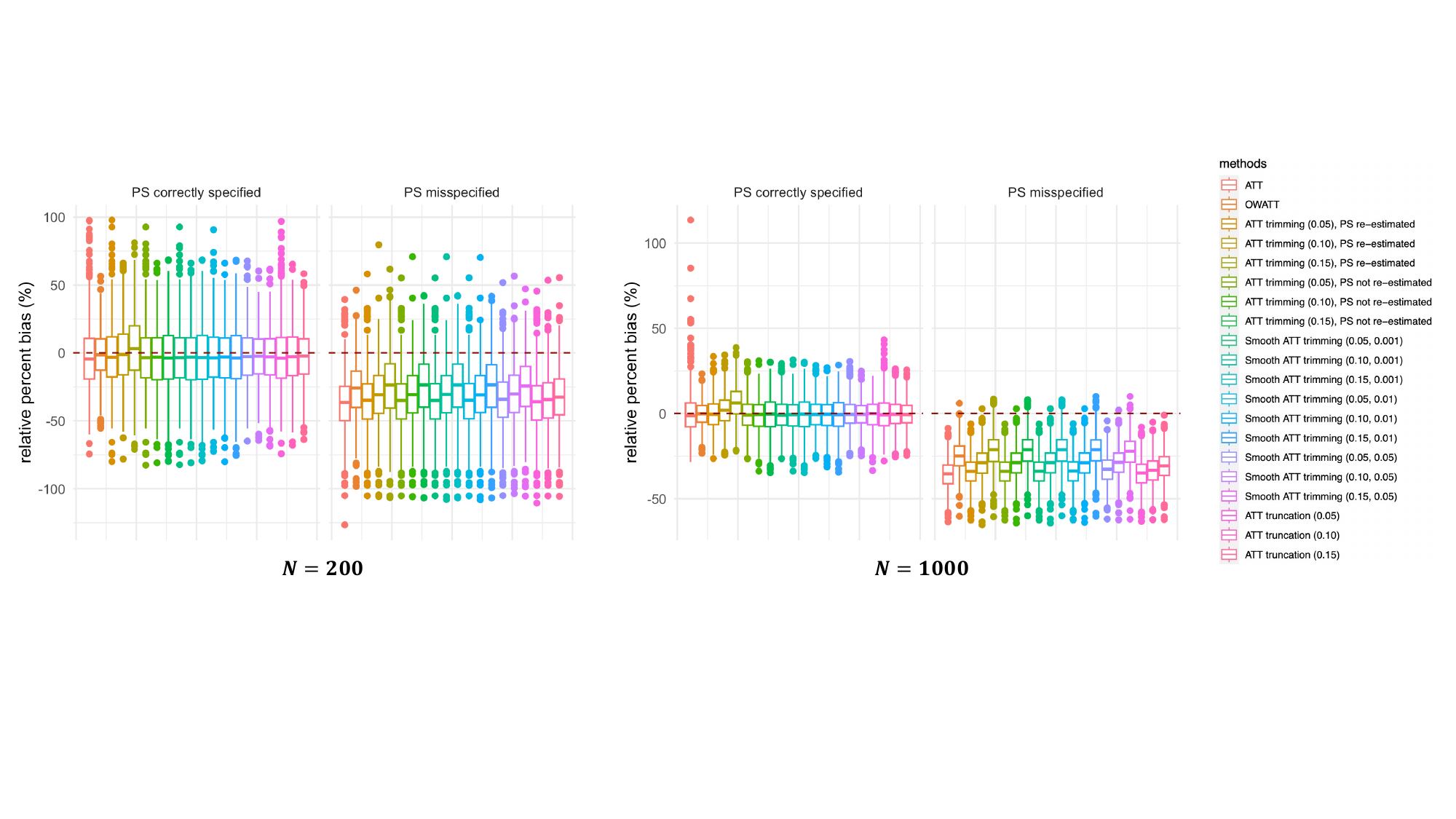}
    \caption{\color{black}Boxplots of relative percent biases by cases under DGP II}
    \label{fig:Rbias-Kang}
\end{figure}

\begin{table}[H]
    \singlespacing
    \scriptsize
    \centering
    \textcolor{black}{
    \begin{tabular}{rcccccccccc}
    \toprule
        Method & ARBias\% & RRMSE & RMSE & RE & CP\% \\
        \midrule
         & \multicolumn{5}{c}{PS model correctly specified} \\
         \cmidrule(rl){2-6}
        ATT & 2.54 & 0.243 & 4.85 & 0.99 & 91.80 \\ 
        \addlinespace
        \bf OWATT & \bf 1.56 & \bf 0.170 & \bf 2.86 & \bf 0.90 & \bf 95.70 \\ 
        \addlinespace
        ATT trimming ($\alpha=0.05$), PS re-estimated & 1.99 & 0.226 & 4.40 & 0.97 & 93.20 \\ 
  ATT trimming ($\alpha=0.10$), PS re-estimated & 0.88 & 0.220 & 3.91 & 0.87 & 96.40 \\ 
  ATT trimming ($\alpha=0.15$), PS re-estimated & 3.10 & 0.240 & 3.74 & 0.81 & 95.90 \\ 
        \addlinespace
     ATT trimming ($\alpha=0.05$), PS not re-estimated & 2.47 & 0.224 & 4.36 & 0.92 & 93.50 \\ 
  ATT trimming ($\alpha=0.10$), PS not re-estimated & 3.77 & 0.226 & 4.02 & 0.83 & 96.50 \\ 
  ATT trimming ($\alpha=0.15$), PS not re-estimated & 4.05 & 0.249 & 3.87 & 0.77 & 96.40 \\  
         \addlinespace
        Smooth ATT trimming ($\alpha=0.05, \varepsilon=0.001$) & 2.48 & 0.223 & 4.34 & 0.93 & 93.40 \\ 
  Smooth ATT trimming ($\alpha=0.10, \varepsilon=0.001$) & 3.72 & 0.225 & 3.99 & 0.83 & 96.70 \\ 
  Smooth ATT trimming ($\alpha=0.15, \varepsilon=0.001$) & 4.00 & 0.248 & 3.86 & 0.78 & 96.50 \\ 
        \addlinespace
        Smooth ATT trimming ($\alpha=0.05, \varepsilon=0.01$) & 2.46 & 0.217 & 4.22 & 0.98 & 93.00 \\ 
  Smooth ATT trimming ($\alpha=0.10, \varepsilon=0.01$) & 3.54 & 0.215 & 3.81 & 0.84 & 96.20 \\ 
  Smooth ATT trimming ($\alpha=0.15, \varepsilon=0.01$) & 3.88 & 0.240 & 3.73 & 0.79 & 96.40 \\ 
        \addlinespace
        Smooth ATT trimming ($\alpha=0.05, \varepsilon=0.05$) & 2.32 & 0.203 & 3.86 & 1.02 & 93.40 \\ 
  Smooth ATT trimming ($\alpha=0.10, \varepsilon=0.05$) & 2.65 & 0.187 & 3.30 & 0.94 & 96.00 \\ 
  Smooth ATT trimming ($\alpha=0.15, \varepsilon=0.05$) & 3.17 & 0.205 & 3.20 & 0.84 & 94.80 \\ 
        \addlinespace
        ATT truncation ($\alpha=0.05$) & 2.41 & 0.232 & 4.60 & 1.08 & 91.60 \\ 
  ATT truncation ($\alpha=0.10$) & 2.41 & 0.207 & 3.97 & 1.04 & 92.90 \\ 
  ATT truncation ($\alpha=0.15$) & 2.46 & 0.191 & 3.50 & 0.99 & 94.60 \\ 
        \addlinespace
        \midrule & \multicolumn{5}{c}{PS model misspecified}\\
         \cmidrule(rl){2-6}
       ATT & 36.72 & 0.420 & 8.39 & 0.91 & 50.80 \\ 
       \addlinespace
       \bf OWATT & \bf 26.42 & \bf 0.333 & \bf 5.61 &\bf 0.93 & \bf 76.30 \\  
        \addlinespace
        ATT trimming ($\alpha=0.05$), PS re-estimated & 35.29 & 0.408 & 7.94 & 0.95 & 55.80 \\ 
  ATT trimming ($\alpha=0.10$), PS re-estimated & 30.53 & 0.372 & 6.61 & 0.88 & 69.70 \\ 
  ATT trimming ($\alpha=0.15$), PS re-estimated & 24.22 & 0.341 & 5.30 & 0.83 & 85.90 \\ 
        \addlinespace
     ATT trimming ($\alpha=0.05$), PS not re-estimated & 35.26 & 0.408 & 7.93 & 0.96 & 55.70 \\ 
  ATT trimming ($\alpha=0.10$), PS not re-estimated & 30.53 & 0.370 & 6.58 & 0.88 & 69.50 \\ 
  ATT trimming ($\alpha=0.15$), PS not re-estimated & 24.21 & 0.337 & 5.25 & 0.84 & 85.20 \\ 
         \addlinespace
        Smooth ATT trimming ($\alpha=0.05, \varepsilon=0.001$) & 35.25 & 0.407 & 7.92 & 0.96 & 55.80 \\ 
  Smooth ATT trimming ($\alpha=0.10, \varepsilon=0.001$) & 30.51 & 0.370 & 6.57 & 0.88 & 69.40 \\ 
  Smooth ATT trimming ($\alpha=0.15, \varepsilon=0.001$) & 24.20 & 0.337 & 5.24 & 0.84 & 84.90 \\ 
        \addlinespace
        Smooth ATT trimming ($\alpha=0.05, \varepsilon=0.01$) & 35.13 & 0.405 & 7.88 & 0.97 & 55.40 \\ 
  Smooth ATT trimming ($\alpha=0.10, \varepsilon=0.01$) & 30.55 & 0.370 & 6.56 & 0.90 & 68.70 \\ 
  Smooth ATT trimming ($\alpha=0.15, \varepsilon=0.01$) & 24.12 & 0.334 & 5.20 & 0.85 & 84.40 \\ 
        \addlinespace
        Smooth ATT trimming ($\alpha=0.05, \varepsilon=0.05$) & 34.16 & 0.396 & 7.53 & 0.98 & 57.70 \\ 
  Smooth ATT trimming ($\alpha=0.10, \varepsilon=0.05$) & 30.38 & 0.366 & 6.44 & 0.95 & 67.50 \\ 
  Smooth ATT trimming ($\alpha=0.15, \varepsilon=0.05$) & 24.60 & 0.330 & 5.16 & 0.90 & 82.70 \\ 
        \addlinespace
        ATT truncation ($\alpha=0.05$) & 36.24 & 0.415 & 8.24 & 0.98 & 51.80 \\ 
  ATT truncation ($\alpha=0.10$) & 34.63 & 0.400 & 7.70 & 0.98 & 56.40 \\ 
  ATT truncation ($\alpha=0.15$) & 32.33 & 0.381 & 6.98 & 0.97 & 62.10 \\
       \bottomrule
    \end{tabular} }
    \begin{tablenotes}\scriptsize
        \item ATT: average treatment effect on the treated; OWATT: overlap weighted average treatment effect on the treated; PS: propensity score; ARBias\%: absolute relative percent bias; RRMSE: relative root mean square error; RMSE: root mean square error; RE: relative efficiency; CP\%: coverage probability (\%). 
    \end{tablenotes}
    \caption{Simulation results using Kang and Schafer's model, $N=200$}\label{tab:res-kang-n200}
\end{table}

\begin{table}[H]
    \singlespacing
    \scriptsize
    \centering
    \textcolor{black}{
    \begin{tabular}{rccccc}
    \toprule
        Method & ARBias\% & RRMSE & RMSE & RE & CP\% \\
        \midrule
         & \multicolumn{5}{c}{PS model correctly specified} \\
         \cmidrule(rl){2-6}
        ATT & 0.12 & 0.125 & 2.50 & 1.10 & 92.80 \\
        \addlinespace
        \bf OWATT & \bf 0.31 & \bf 0.073 & \bf 1.22 & \bf 0.92 & \bf 95.30 \\ 
        \addlinespace
        ATT trimming ($\alpha=0.05$), PS re-estimated & 0.31 & 0.093 & 1.80 & 0.84 & 94.10 \\ 
  ATT trimming ($\alpha=0.10$), PS re-estimated & 1.67 & 0.092 & 1.64 & 0.87 & 95.20 \\ 
  ATT trimming ($\alpha=0.15$), PS re-estimated & 6.30 & 0.117 & 1.82 & 0.84 & 91.40 \\
        \addlinespace
     ATT trimming ($\alpha=0.05$), PS not re-estimated & 0.76 & 0.093 & 1.81 & 0.81 & 94.20 \\ 
  ATT trimming ($\alpha=0.10$), PS not re-estimated & 1.00 & 0.094 & 1.68 & 0.85 & 95.70 \\ 
  ATT trimming ($\alpha=0.15$), PS not re-estimated & 0.57 & 0.105 & 1.63 & 0.81 & 96.50 \\ 
         \addlinespace
        Smooth ATT trimming ($\alpha=0.05, \varepsilon=0.001$) & 0.77 & 0.092 & 1.80 & 0.82 & 94.10 \\ 
  Smooth ATT trimming ($\alpha=0.10, \varepsilon=0.001$) & 0.97 & 0.094 & 1.66 & 0.85 & 95.60 \\ 
  Smooth ATT trimming ($\alpha=0.15, \varepsilon=0.001$) & 0.60 & 0.104 & 1.62 & 0.82 & 96.30 \\ 
        \addlinespace
        Smooth ATT trimming ($\alpha=0.05, \varepsilon=0.01$) & 0.59 & 0.089 & 1.73 & 0.86 & 94.10 \\ 
  Smooth ATT trimming ($\alpha=0.10, \varepsilon=0.01$) & 0.91 & 0.090 & 1.59 & 0.88 & 95.00 \\ 
  Smooth ATT trimming ($\alpha=0.15, \varepsilon=0.01$) & 0.67 & 0.101 & 1.57 & 0.84 & 95.80 \\ 
        \addlinespace
        Smooth ATT trimming ($\alpha=0.05, \varepsilon=0.05$) & 0.31 & 0.085 & 1.62 & 0.94 & 94.20 \\ 
  Smooth ATT trimming ($\alpha=0.10, \varepsilon=0.05$) & 0.57 & 0.077 & 1.36 & 0.90 & 95.60 \\ 
  Smooth ATT trimming ($\alpha=0.15, \varepsilon=0.05$) & 0.64 & 0.088 & 1.37 & 0.91 & 95.60 \\ 
        \addlinespace
        ATT truncation ($\alpha=0.05$) & 0.18 & 0.102 & 2.02 & 0.98 & 93.30 \\ 
  ATT truncation ($\alpha=0.10$) & 0.42 & 0.086 & 1.65 & 0.92 & 94.30 \\ 
  ATT truncation ($\alpha=0.15$) & 0.48 & 0.078 & 1.43 & 0.90 & 95.10 \\ 
        \addlinespace
        \midrule & \multicolumn{5}{c}{PS model misspecified}\\
         \cmidrule(rl){2-6}
       ATT & 35.63 & 0.366 & 7.32 & 0.92 &  3.40 \\  
       \addlinespace
       \bf OWATT & \bf 24.83 & \bf 0.263 & \bf 4.43 & \bf 0.90 & \bf 20.10 \\ 
        \addlinespace
        ATT trimming ($\alpha=0.05$), PS re-estimated & 34.09 & 0.351 & 6.84 & 0.91 &  3.90 \\ 
  ATT trimming ($\alpha=0.10$), PS re-estimated & 28.90 & 0.303 & 5.37 & 0.87 & 13.60 \\ 
  ATT trimming ($\alpha=0.15$), PS re-estimated & 21.73 & 0.240 & 3.73 & 0.86 & 49.40 \\
        \addlinespace
     ATT trimming ($\alpha=0.05$), PS not re-estimated & 34.09 & 0.351 & 6.84 & 0.90 &  3.80 \\ 
  ATT trimming ($\alpha=0.10$), PS not re-estimated & 28.90 & 0.302 & 5.37 & 0.86 & 12.80 \\ 
  ATT trimming ($\alpha=0.15$), PS not re-estimated & 21.55 & 0.237 & 3.69 & 0.84 & 48.40 \\ 
         \addlinespace
        Smooth ATT trimming ($\alpha=0.05, \varepsilon=0.001$) & 34.10 & 0.351 & 6.84 & 0.91 &  3.90 \\ 
  Smooth ATT trimming ($\alpha=0.10, \varepsilon=0.001$) & 28.90 & 0.302 & 5.37 & 0.86 & 12.50 \\ 
  Smooth ATT trimming ($\alpha=0.15, \varepsilon=0.001$) & 21.55 & 0.237 & 3.69 & 0.85 & 48.20 \\ 
        \addlinespace
        Smooth ATT trimming ($\alpha=0.05, \varepsilon=0.01$) & 34.00 & 0.350 & 6.81 & 0.91 &  3.60 \\ 
  Smooth ATT trimming ($\alpha=0.10, \varepsilon=0.01$) & 28.91 & 0.302 & 5.36 & 0.88 & 11.50 \\ 
  Smooth ATT trimming ($\alpha=0.15, \varepsilon=0.01$) & 21.66 & 0.238 & 3.70 & 0.87 & 46.40 \\
        \addlinespace
        Smooth ATT trimming ($\alpha=0.05, \varepsilon=0.05$) & 32.90 & 0.340 & 6.46 & 0.92 &  3.40 \\ 
  Smooth ATT trimming ($\alpha=0.10, \varepsilon=0.05$) & 28.79 & 0.301 & 5.29 & 0.90 & 10.40 \\ 
  Smooth ATT trimming ($\alpha=0.15, \varepsilon=0.05$) & 22.55 & 0.244 & 3.82 & 0.89 & 36.90 \\ 
        \addlinespace
        ATT truncation ($\alpha=0.05$) & 35.17 & 0.362 & 7.18 & 0.93 &  3.00 \\ 
  ATT truncation ($\alpha=0.10$) & 33.41 & 0.345 & 6.63 & 0.92 &  3.60 \\ 
  ATT truncation ($\alpha=0.15$) & 30.87 & 0.320 & 5.87 & 0.91 &  6.20 \\ 
       \bottomrule
    \end{tabular}}
    \begin{tablenotes}\scriptsize
        \item ATT: average treatment effect on the treated; OWATT: overlap weighted average treatment effect on the treated; PS: propensity score; ARBias\%: absolute relative percent bias; RRMSE: relative root mean square error; RMSE: root mean square error; RE: relative efficiency; CP\%: coverage probability (\%). 
    \end{tablenotes}
    \caption{Simulation results using Kang and Schafer's model, $N=1000$}\label{tab:res-kang-n1000}
\end{table}

\section{Appendix: Additional Data Analysis Results}\label{apx:datana}
In this section, we show an additional detail of the MEPS data analysis in Section \ref{sec:data}. The following love plot in Figure \ref{fig:love-meps} displays the {absolute standardized mean differences (ASD) }\citep{greifer2020covariate,austin2015moving} of 31 covariates between White and Asian group using weights in all 20 methods we considered in both simulation and data analysis. We use different shapes of points to indicate different thresholds we select within a given ad-hoc method. 

As can be seen, {ASDs by ATT weights exceed the $0.1$ threshold frequently and sometimes in a larger degree than other methods}, and overall, OWATT, ATT trimming ($\alpha=0.05$) with PS re-estimated, Smooth ATT trimming ($\alpha=0.05, \varepsilon=0.001$), Smooth ATT trimming ($\alpha=0.05, \varepsilon=0.01$), Smooth ATT trimming ($\alpha=0.05, \varepsilon=0.05$), Smooth ATT trimming ($\alpha=0.10, \varepsilon=0.05$), ATT truncation ($\alpha=0.05$), ATT truncation ($\alpha=0.10$) balance covariates better than other methods.

\begin{figure}[H]
    \centering  \includegraphics[width=1\textwidth]{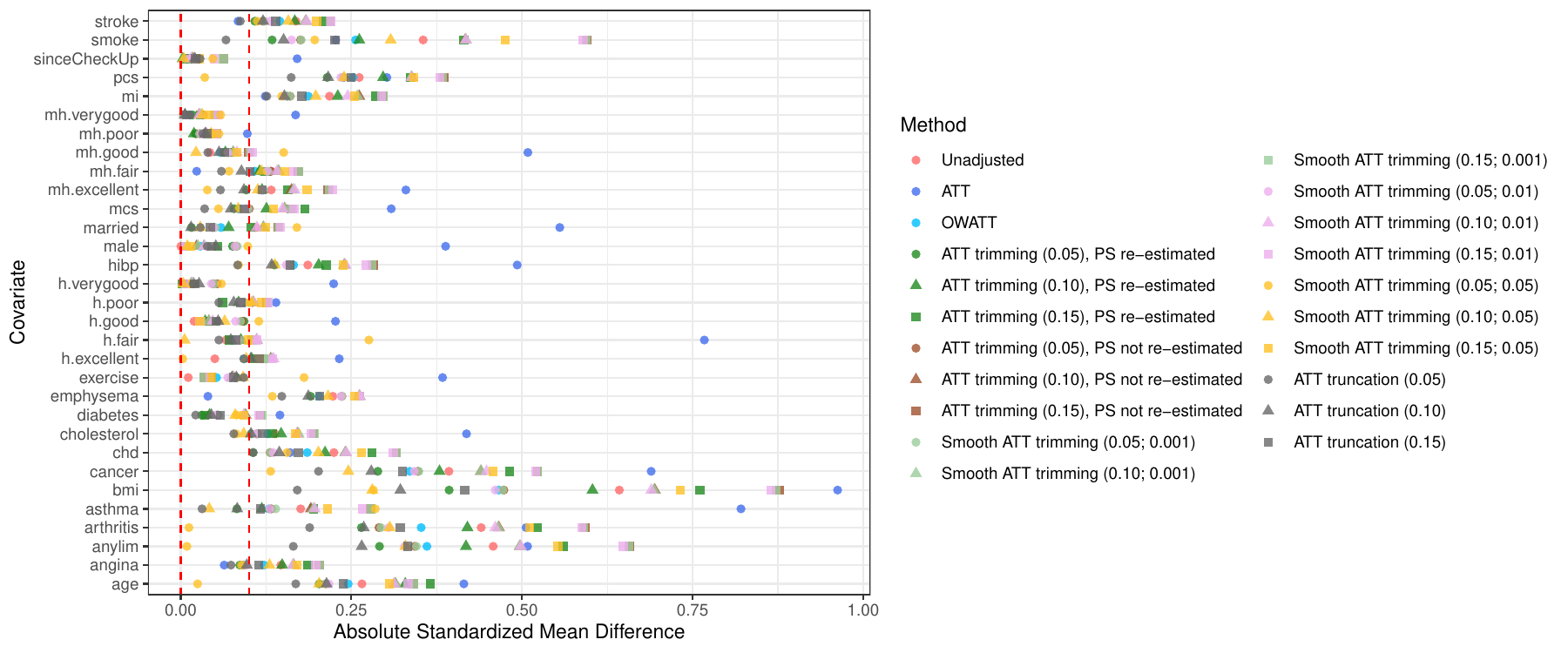}   
    \begin{tablenotes}\small
        \item The two red vertical dashed lines indicate the absolute difference within $\pm 0.10$. 
    \end{tablenotes}
    \caption{Covariates balance of White vs. Asian in MEPS data}
    \label{fig:love-meps}
\end{figure}

\end{document}